%% file: 00_Article_Merge.tex
\documentclass[times, twoside]{zHenriquesLab-StyleBioRxiv}
\usepackage{blindtext}
\usepackage[overload]{textcase}
\usepackage{pdfpages}

\begin{document}
    \bibliographystyle{zHenriquesLab-StyleBib}
    \include{01_Article_MainText}
    

\section*{Supplementary material}

Table of Content

# Supplementary Methods

**Supplementary Methods 1. PubMed search terms**

To get a rough estimate of the number of studies specifically investigating case-control differences between healthy and depressive subjects in neuroimaging modalities, we conducted a PubMed (https://pubmed.ncbi.nlm.nih.gov/advanced/) search on the 28th of September 2021 using the following search term which includes depression, a healthy control group and neuroimaging as key search components, resulting in a total of 1,585 publications:

*(((major depressive disorder[Title]) OR (depression[Title]) OR (major depression[Title])) AND ((neuroimaging[Title/Abstract]) OR (magnetic resonance imaging[Title/Abstract]) OR (MRI[Title/Abstract])) AND ((healthy[Title/Abstract]) OR (control group[Title/Abstract]) OR (case control[Title/Abstract])))*

**Supplementary Methods 2. Remission status**

MDD diagnosis was based on the criteria defined in DSM-IV which requires the presence of five of nine cardinal symptoms for two weeks or longer within the last 4 weeks, are present for most of the day nearly every day and need to cause significant distress or impairment. To fulfill the diagnostic criteria, a depressed mood or markedly diminished interest or pleasure have to be present (at least one or both). Other symptoms are clinically significant weight gain/loss or appetite disturbance, insomnia or hypersomnia, psychomotor agitation or retardation, fatigue or loss of energy, feelings of worthlessness or excessive guilt, diminished ability to concentrate or think clearly, and recurrent thoughts of death or suicide.

Partial remission was defined either by (1) a presence of some major depressive symptoms but full criteria are no longer met or (2) no major depressive symptoms but the period of remission has been less than two months. Complete remission was defined as the disappearance of the diagnostic criteria of depression for at least two consecutive months.

**Supplementary Methods 3. Polygenic Risk Scores**

Genotyping was conducted using the PsychArray BeadChip, followed by quality control and imputation, as described previously.[1,2] QC and population substructure

analyses were performed using PLINK v1.90.[3] The data were imputed to the 1000 Genomes phase 3 reference panel using SHAPEIT and IMPUTE2.[4,5] When calculating polygenic risk scores (PRS), SNP weights were estimated using the PRS-CS method with default parameters.[6,7] This method employs Bayesian regression to infer PRS weights while modeling the local linkage disequilibrium patterns of all SNPs using the EUR super-population of the 1000 Genomes reference panel. Using these weights, the PRS were calculated in PLINK v1.90 on imputed dosage data.

A Polygenic Risk Score (PRS) for Major Depression (MDD) was used in the genetic analyses.[8] The global shrinkage parameter $\varphi$ was automatically determined (PRS-CS-auto; $\varphi=1.30\times10^{-4}$). For the calculation of ancestry components, the pairwise identity-by-state matrix of all individuals was calculated on strictly quality-controlled pre-imputation genotype data. Multidimensional scaling (MDS) analysis was performed on this matrix using the eigendecomposition-based algorithm in PLINK v1.90. Prior to the calculation of MDS components, related subjects were identified and excluded from the sample, keeping only one of the siblings in the sample. This procedure was repeated whenever the analyzed sample changed, e.g., when comparing only acutely depressed patients with healthy subjects.

**Supplementary Methods 4. Functional MRI image acquisition**

For the two functional MRI paradigms (face matching, resting-state), a T2*-weighted echo-planar imaging (EPI) sequence sensitive to blood oxygen level-dependent (BOLD) contrast was used with the following parameters: TE = 30 ms (Marburg), TE = 29ms (Münster), TR = 2,000 ms, FoV = 210 mm, matrix = 64 × 64, slice thickness = 3.8 mm, distance factor = 10%, phase encoding direction anterior >> posterior, flip angle = 90°, no parallel imaging, bandwidth 2,232 Hz/Px, ascending acquisition, axial acquisition, 33 slices, slice alignment parallel to AC-PC line tilted 20° in the dorsal direction.

For resting-state fMRI, two-hundred-thirty-seven interleaved and ascending measurements (8 minutes) tilted with -20° against the anterior and posterior commission alignment (AC-PC alignment) were acquired (Base resolution: 64, Bandwith: 2232 Hz/Px, Echo spacing: 0.51ms EPI factor: 64, BOLD threshold: 4, TR:

2s). Participants were asked to keep their eyes closed until the end of the resting state session.

**Supplementary Methods 5. DTI image acquisition and preprocessing**

*Preprocessing*

First, FSL's eddy was used to realign the Diffusion-weighted images (DWI) and correct those for eddy currents and susceptibility distortions.[9–12] Second, the CATO toolbox was used to reconstruct the anatomical connectome of the diffusion tensor imaging data (DTI) which models the measured signal of a single voxel by a tensor describing the preferred diffusion direction per voxel. It makes use of the RESTORE algorithm which estimates the diffusion tensor while simultaneously identifying and removing outliers, thereby reducing the impact of physiological noise artifacts.[10,11] Third, the resulting diffusion profiles were used to reconstruct white matter paths using deterministic tractography. To this end, eight seeds were started per voxel, and for each seed, a tractography streamline was constructed by following the main diffusion direction from voxel to voxel. Stop criteria included reaching a voxel with a fractional anisotropy < 0.1, making a sharp turn of >45°, reaching a gray matter voxel, or exiting the brain mask [16]. Given the poorer DWI signal-to-noise ratio in subcortical regions and the dominant effect of subcortical regions on network properties, we decided to use a subdivision of FreeSurfer's Desikan Killiany Atlas containing only cortical regions, as we have done in previous work.[13–16]

*Quality control*

Four metrics were included in the detection of outliers: the average number of streamlines, the average fractional anisotropy, the average prevalence of each subject's connections (low value if the subject has "odd" connections), and the average prevalence of each subjects connected brain regions (high value, if the subject misses commonly found connections). Then, quartiles (Q1, Q2, Q3) and the interquartile range (IQR=Q3-Q1) was computed for every metric across the group. A datapoint was declared an outlier if its value was below Q1-1.5*IQR or above Q3+1.5*IQR on any of the four metrics.

**Supplementary Methods 6. Freesurfer measures**

L_bankssts_thickavg
L_caudalanteriorcingulate_thickavg
L_caudalmiddlefrontal_thickavg
L_cuneus_thickavg
L_entorhinal_thickavg
L_fusiform_thickavg
L_inferiorparietal_thickavg
L_inferiortemporal_thickavg
L_isthmuscingulate_thickavg
L_lateraloccipital_thickavg
L_lateralorbitofrontal_thickavg
L_lingual_thickavg
L_medialorbitofrontal_thickavg
L_middletemporal_thickavg
L_parahippocampal_thickavg
L_paracentral_thickavg
L_parsopercularis_thickavg
L_parsorbitalis_thickavg
L_parstriangularis_thickavg
L_pericalcarine_thickavg
L_postcentral_thickavg
L_posteriorcingulate_thickavg
L_precentral_thickavg
L_precuneus_thickavg
L_rostralanteriorcingulate_thickavg
L_rostralmiddlefrontal_thickavg
L_superiorfrontal_thickavg
L_superiorparietal_thickavg
L_superiortemporal_thickavg
L_supramarginal_thickavg
L_frontalpole_thickavg
L_temporalpole_thickavg
L_transversetemporal_thickavg
L_insula_thickavg
R_bankssts_thickavg
R_caudalanteriorcingulate_thickavg
R_caudalmiddlefrontal_thickavg
R_cuneus_thickavg
R_entorhinal_thickavg
R_fusiform_thickavg
R_inferiorparietal_thickavg
R_inferiortemporal_thickavg
R_isthmuscingulate_thickavg
R_lateraloccipital_thickavg
R_lateralorbitofrontal_thickavg

R_lingual_thickavg
R_medialorbitofrontal_thickavg
R_middletemporal_thickavg
R_parahippocampal_thickavg
R_paracentral_thickavg
R_parsopercularis_thickavg
R_parsorbitalis_thickavg
R_parstriangularis_thickavg
R_pericalcarine_thickavg
R_postcentral_thickavg
R_posteriorcingulate_thickavg
R_precentral_thickavg
R_precuneus_thickavg
R_rostralanteriorcingulate_thickavg
R_rostralmiddlefrontal_thickavg
R_superiorfrontal_thickavg
R_superiorparietal_thickavg
R_superiortemporal_thickavg
R_supramarginal_thickavg
R_frontalpole_thickavg
R_temporalpole_thickavg
R_transversetemporal_thickavg
R_insula_thickavg
LThickness
RThickness
LSurfArea
RSurfArea
L_bankssts_surfavg
L_caudalanteriorcingulate_surfavg
L_caudalmiddlefrontal_surfavg
L_cuneus_surfavg
L_entorhinal_surfavg
L_fusiform_surfavg
L_inferiorparietal_surfavg
L_inferiortemporal_surfavg
L_isthmuscingulate_surfavg
L_lateraloccipital_surfavg
L_lateralorbitofrontal_surfavg
L_lingual_surfavg
L_medialorbitofrontal_surfavg
L_middletemporal_surfavg
L_parahippocampal_surfavg
L_paracentral_surfavg
L_parsopercularis_surfavg
L_parsorbitalis_surfavg
L_parstriangularis_surfavg
L_pericalcarine_surfavg
L_postcentral_surfavg
L_posteriorcingulate_surfavg

L_precentral_surfavg
L_precuneus_surfavg
L_rostralanteriorcingulate_surfavg
L_rostralmiddlefrontal_surfavg
L_superiorfrontal_surfavg
L_superiorparietal_surfavg
L_superiortemporal_surfavg
L_supramarginal_surfavg
L_frontalpole_surfavg
L_temporalpole_surfavg
L_transversetemporal_surfavg
L_insula_surfavg
R_bankssts_surfavg
R_caudalanteriorcingulate_surfavg
R_caudalmiddlefrontal_surfavg
R_cuneus_surfavg
R_entorhinal_surfavg
R_fusiform_surfavg
R_inferiorparietal_surfavg
R_inferiortemporal_surfavg
R_isthmuscingulate_surfavg
R_lateraloccipital_surfavg
R_lateralorbitofrontal_surfavg
R_lingual_surfavg
R_medialorbitofrontal_surfavg
R_middletemporal_surfavg
R_parahippocampal_surfavg
R_paracentral_surfavg
R_parsopercularis_surfavg
R_parsorbitalis_surfavg
R_parstriangularis_surfavg
R_pericalcarine_surfavg
R_postcentral_surfavg
R_posteriorcingulate_surfavg
R_precentral_surfavg
R_precuneus_surfavg
R_rostralanteriorcingulate_surfavg
R_rostralmiddlefrontal_surfavg
R_superiorfrontal_surfavg
R_superiorparietal_surfavg
R_superiortemporal_surfavg
R_supramarginal_surfavg
R_frontalpole_surfavg
R_temporalpole_surfavg
R_transversetemporal_surfavg
R_insula_surfavg
ICV
LLatVent
RLatVent

Lthal
Rthal
Lcaud
Rcaud
Lput
Rput
Lpal
Rpal
Lhippo
Rhippo
Lamyg
Ramyg
Laccumb
Raccumb
BrainSegVol
BrainSegVolNotVent
lhCortexVol
rhCortexVol
CortexVol
lh.WhiteMatterVol
rh.WhiteMatterVol
SubCortGrayVol
TotalGrayVol

# Supplementary Results

## Supplementary Table S1. Results for males and females

Supplementary Table S1

*Statistical analyses of differences between depressive and healthy subjects analyzed separately for males and females*

| **HC versus MDD (male)** | n | df1 | df2 | F | $p_{uncorr}$ | $p_{corr}$ | $h^2_{partial}$ | overlap | BACC |
|---|---|---|---|---|---|---|---|---|---|
| Structural MRI | | | | | | | | | |
|   Cortical and subcortical surface, thickness, volume | 611 | 1 | 606 | 9.94 | 0.002 | 1.000 | 0.016 [0.003 - 0.042] | 89.91% | 55.49% |
| Task-based functional MRI | | | | | | | | | |
|   Face matching task | 425 | 1 | 420 | 8.56 | 0.004 | 0.557 | 0.020 [0.002 - 0.052] | 84.89% | 52.94% |
| Structural connectome | | | | | | | | | |
|   FA | 530 | 1 | 525 | 6.33 | 0.012 | 1.000 | 0.012 [0.001 - 0.034] | 89.38% | 56.60% |
|   MD | 528 | 1 | 523 | 12.83 | $3.73 \times 10^{-4}$ | 0.543 | 0.024 [0.005 - 0.055] | 84.16% | 57.25% |
|   Network parameter | 533 | 1 | 528 | 9.05 | 0.003 | 1.000 | 0.017 [0.002 - 0.043] | 89.73% | 55.65% |
| Functional connectome | | | | | | | | | |
|   Bivariate connectivity | 462 | 1 | 457 | 17.47 | $3.50 \times 10^{-5}$ | 0.457 | 0.037 [0.011 - 0.075] | 85.01% | 55.43% |
|   Network parameter | 462 | 1 | 457 | 16.13 | $6.92 \times 10^{-5}$ | 0.111 | 0.034 [0.008 - 0.073] | 84.77% | 56.79% |
| Genetics | | | | | | | | | |
|   Polygenic Risk Scores | 591 | 1 | 584 | 25.86 | $4.96 \times 10^{-7}$ | $4.96 \times 10^{-7}$ | 0.042 [0.017 - 0.079] | 83.60% | 58.49% |
| Environment | | | | | | | | | |
|   Social Support | 640 | 1 | 636 | 178.80 | $4.09 \times 10^{-36}$ | $4.09 \times 10^{-36}$ | 0.219 [0.158 - 0.276] | 58.78% | 71.83% |
|   Childhood Maltreatment | 640 | 1 | 636 | 121.62 | $5.36 \times 10^{-26}$ | $5.36 \times 10^{-26}$ | 0.161 [0.111 - 0.209] | 60.12% | 66.83% |
| **HC versus MDD (female)** | | | | | | | | | |
| Structural MRI | | | | | | | | | |
|   Cortical and subcortical surface, thickness, volume | 1130 | 1 | 1125 | 14.07 | $1.85 \times 10^{-4}$ | 0.090 | 0.012 [0.003 - 0.027] | 91.20% | 53.91% |
| Task-based functional MRI | | | | | | | | | |
|   Face matching task | 816 | 1 | 811 | 14.95 | $1.19 \times 10^{-4}$ | 0.628 | 0.018 [0.004 - 0.041] | 89.56% | 55.57% |
| Structural connectome | | | | | | | | | |
|   FA | 975 | 1 | 970 | 10.88 | 0.001 | 1.000 | 0.011 [0.002 - 0.029] | 91.51% | 54.65% |
|   MD | 939 | 1 | 934 | 15.17 | $1.05 \times 10^{-4}$ | 0.144 | 0.016 [0.003 - 0.033] | 85.22% | 52.86% |
|   Network parameter | 975 | 1 | 970 | 9.33 | 0.002 | 1.000 | 0.010 [0.001 - 0.025] | 91.62% | 54.02% |
| Functional connectome | | | | | | | | | |
|   Bivariate connectivity | 883 | 1 | 878 | 18.51 | $1.88 \times 10^{-5}$ | 0.791 | 0.021 [0.007 - 0.047] | 88.22% | 54.64% |
|   Network parameter | 883 | 1 | 878 | 9.22 | 0.002 | 1.000 | 0.010 [0.001 - 0.030] | 92.10% | 53.83% |
| Genetics | | | | | | | | | |
|   Polygenic Risk Scores | 1053 | 1 | 1046 | 25.99 | $4.06 \times 10^{-7}$ | $4.06 \times 10^{-7}$ | 0.024 [0.009 - 0.045] | 87.56% | 57.92% |
| Environment | | | | | | | | | |
|   Social Support | 1157 | 1 | 1153 | 302.90 | $2.04 \times 10^{-60}$ | $2.04 \times 10^{-60}$ | 0.208 [0.170 - 0.245] | 54.29% | 69.89% |
|   Childhood Maltreatment | 1159 | 1 | 1155 | 306.47 | $4.82 \times 10^{-61}$ | $4.82 \times 10^{-61}$ | 0.210 [0.175 - 0.246] | 53.33% | 72.40% |

BACC = balanced accuracy, HC = Healthy Controls, MDD = Major Depressive Disorder, FA = fractional anisotropy, MD = mean diffusivity

# Supplementary Table S2. Results for Marburg and Münster

Supplementary Table S2
*Statistical analyses of differences between depressive and healthy subjects analyzed separately for Marburg and Münster*

| HC versus MDD (Marburg) | n | df1 | df2 | F | $p_{uncorr}$ | $p_{corr}$ | $\eta^2_{partial}$ | overlap | BACC |
|---|---|---|---|---|---|---|---|---|---|
| Structural MRI | | | | | | | | | |
|   Cortical and subcortical surface, thickness, volume | 914 | 1 | 910 | 8.99 | 0.003 | 0.718 | 0.010 [0.001 - 0.026] | 91.86 % | 54.92 % |
| Task-based functional MRI | | | | | | | | | |
|   Face matching task | 806 | 1 | 802 | 13.07 | $3.18 \times 10^{-4}$ | 0.558 | 0.016 [0.003 - 0.034] | 88.98 % | 55.48 % |
| Structural connectome | | | | | | | | | |
|   FA | 778 | 1 | 774 | 11.15 | 0.001 | 1.000 | 0.014 [0.002 - 0.035] | 90.21 % | 55.22 % |
|   MD | 798 | 1 | 794 | 15.60 | $8.54 \times 10^{-5}$ | 0.121 | 0.019 [0.005 - 0.042] | 88.08 % | 55.43 % |
|   Network parameter | 815 | 1 | 811 | 13.01 | $3.29 \times 10^{-4}$ | 0.735 | 0.016 [0.002 - 0.038] | 89.37 % | 55.74 % |
| Functional connectome | | | | | | | | | |
|   Bivariate connectivity | 876 | 1 | 872 | 26.57 | $3.14 \times 10^{-7}$ | 0.014 | 0.030 [0.013 - 0.057] | 85.33 % | 55.78 % |
|   Network parameter | 876 | 1 | 872 | 15.02 | $1.14 \times 10^{-4}$ | 0.221 | 0.017 [0.004 - 0.036] | 89.29 % | 54.92 % |
| Genetics | | | | | | | | | |
|   Polygenic Risk Scores | 867 | 1 | 860 | 46.50 | $1.73 \times 10^{-11}$ | $1.73 \times 10^{-11}$ | 0.051 [0.028 - 0.086] | 81.43 % | 58.95 % |
| Environment | | | | | | | | | |
|   Social Support | 952 | 1 | 948 | 293.58 | $1.54 \times 10^{-57}$ | $1.54 \times 10^{-57}$ | 0.236 [0.185 - 0.285] | 55.35 % | 70.86 % |
|   Childhood Maltreatment | 950 | 1 | 946 | 269.91 | $1.50 \times 10^{-53}$ | $1.50 \times 10^{-53}$ | 0.222 [0.177 - 0.270] | 52.88 % | 72.87 % |
| **HC versus MDD (Münster)** | | | | | | | | | |
| Structural MRI | | | | | | | | | |
|   Cortical and subcortical surface, thickness, volume | 827 | 1 | 823 | 7.84 | 0.005 | 1.000 | 0.009 [0.001 - 0.027] | 92.30 % | 53.74 % |
| Task-based functional MRI | | | | | | | | | |
|   Face matching task | 435 | 1 | 431 | 15.64 | $8.97 \times 10^{-5}$ | 0.332 | 0.035 [0.008 - 0.075] | 84.41 % | 56.22 % |
| Structural connectome | | | | | | | | | |
|   FA | 692 | 1 | 688 | 10.18 | 0.001 | 1.000 | 0.015 [0.002 - 0.037] | 90.52 % | 53.80 % |
|   MD | 690 | 1 | 686 | 14.48 | $1.55 \times 10^{-4}$ | 0.210 | 0.021 [0.006 - 0.044] | 88.42 % | 56.05 % |
|   Network parameter | 693 | 1 | 689 | 7.91 | 0.005 | 1.000 | 0.011 [0.001 - 0.032] | 91.23 % | 54.44 % |
| Functional connectome | | | | | | | | | |
|   Bivariate connectivity | 469 | 1 | 465 | 13.01 | $3.44 \times 10^{-4}$ | 1.000 | 0.027 [0.005 - 0.060] | 85.98 % | 59.11 % |
|   Network parameter | 469 | 1 | 465 | 10.32 | 0.001 | 1.000 | 0.022 [0.003 - 0.058] | 87.72% | 56.75% |
| Genetics | | | | | | | | | |
|   Polygenic Risk Scores | 752 | 1 | 745 | 13.53 | $2.51 \times 10^{-4}$ | $2.51 \times 10^{-4}$ | 0.018 [0.004 - 0.042] | 89.41 % | 57.52 % |
| Environment | | | | | | | | | |
|   Social Support | 845 | 1 | 841 | 193.68 | $8.94 \times 10^{-40}$ | $8.94 \times 10^{-40}$ | 0.187 [0.147 - 0.228] | 55.62 % | 70.21 % |
|   Childhood Maltreatment | 849 | 1 | 845 | 157.94 | $2.49 \times 10^{-33}$ | $2.49 \times 10^{-33}$ | 0.157 [0.122 - 0.195] | 57.73 % | 69.34 % |

BACC = balanced accuracy, HC = Healthy Controls, MDD = Major Depressive Disorder, FA = fractional anisotropy, MD = mean diffusivity

## Supplementary Figure S1. Predictive utility – Freesurfer- healthy versus depressive subjects

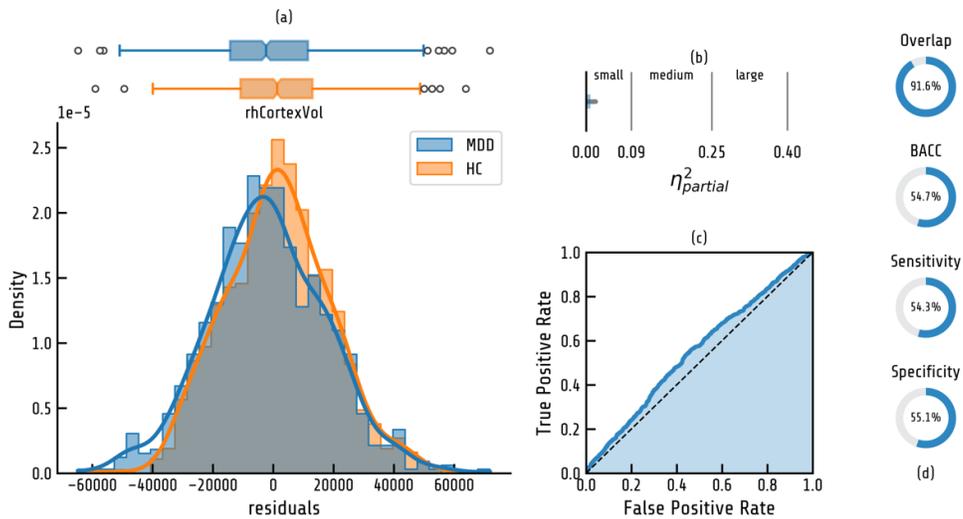

*Figure S 1. Distributional overlap, effect size and classification performance for Freesurfer data. (a) shows a histogram with Gaussian Kernel Density estimation as solid line and boxplot of the confound-corrected values of the Freesurfer region displaying the largest effect. (b) partial $\eta^2$ ANOVA effect size. Light blue indicates upper bound of bootstrapped 95% confidence interval. (c) shows Receiver-Operating-Characteristic (ROC) Curve for Logistic Regression classification. (d) shows distributional overlap, balanced accuracy (BACC) as well as sensitivity and specificity of the Logistic Regression classification.*

## Supplementary Figure S2. Predictive utility – task fMRI - healthy versus depressive subjects

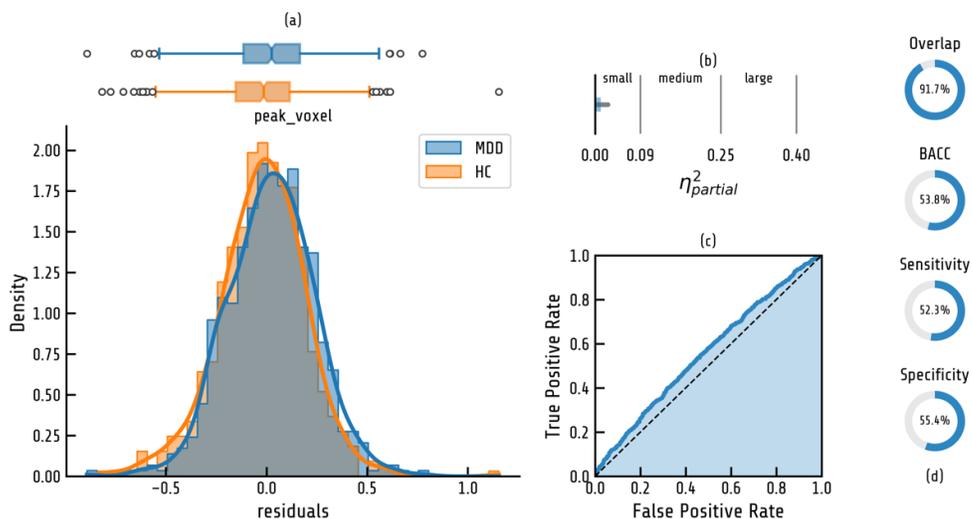

*Figure S 2. Distributional overlap, effect size and classification performance for task-based fMRI data. (a) shows a histogram with Gaussian Kernel Density estimation as solid line and boxplot of the confound-corrected values of the voxel displaying the largest effect. (b) partial $\eta^2$ ANOVA effect size. Light blue indicates upper bound of bootstrapped 95% confidence interval. (c) shows Receiver-Operating-Characteristic (ROC) Curve for Logistic Regression classification. (d) shows distributional overlap, balanced accuracy (BACC) as well as sensitivity and specificity of the Logistic Regression classification.*

## Supplementary Figure S3. Predictive utility – DTI FA - healthy versus depressive subjects

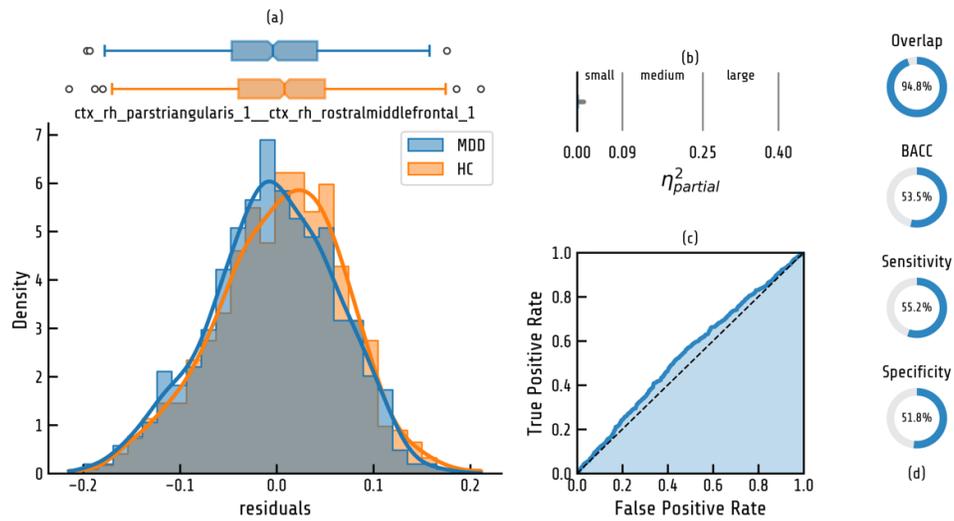

Figure S 3. Distributional overlap, effect size and classification performance for DTI FA. (a) shows a histogram with Gaussian Kernel Density estimation as solid line and boxplot of the confound-corrected values of the connection displaying the largest effect. (b) partial $\eta^2$ ANOVA effect size. Light blue indicates upper bound of bootstrapped 95% confidence interval. (c) shows Receiver-Operating-Characteristic (ROC) Curve for Logistic Regression classification. (d) shows distributional overlap, balanced accuracy (BACC) as well as sensitivity and specificity of the Logistic Regression classification.

## Supplementary Figure S4. Predictive utility – DTI MD - healthy versus depressive subjects

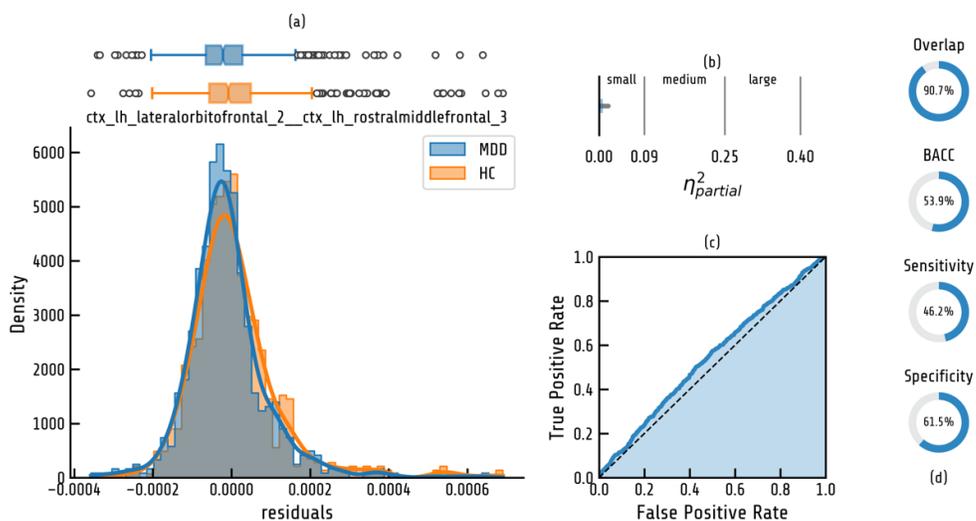

Figure S 4. Distributional overlap, effect size and classification performance for DTI MD. (a) shows a histogram with Gaussian Kernel Density estimation as solid line and boxplot of the confound-corrected values of the connection displaying the largest effect. (b) partial $\eta^2$ ANOVA effect size. Light blue indicates upper bound of bootstrapped 95% confidence interval. (c) shows Receiver-Operating-Characteristic (ROC) Curve for Logistic Regression classification.

*(d) shows distributional overlap, balanced accuracy (BACC) as well as sensitivity and specificity of the Logistic Regression classification.*

## Supplementary Figure S5. Predictive utility – DTI graph - healthy versus depressive subjects

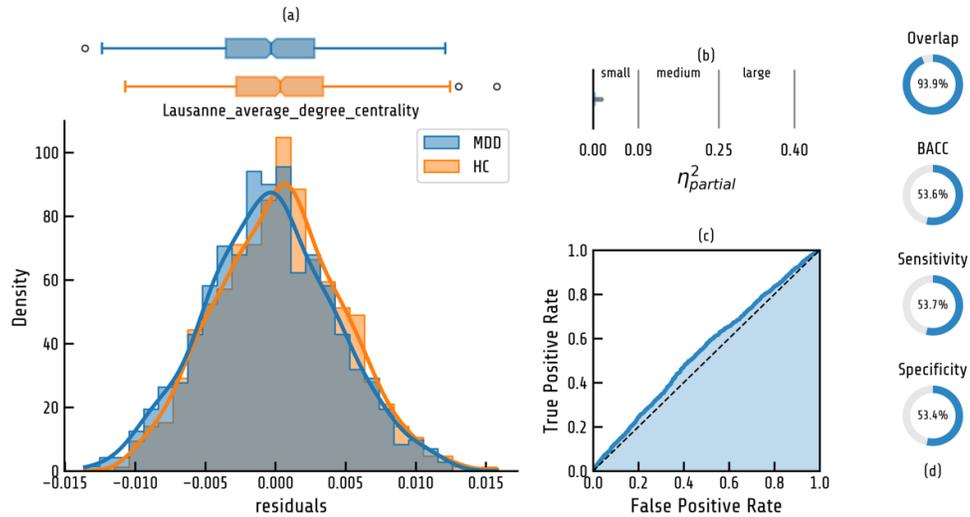

*Figure S 5. Distributional overlap, effect size and classification performance for DTI graph metrics. (a) shows a histogram with Gaussian Kernel Density estimation as solid line and boxplot of the confound-corrected values of the network graph parameter displaying the largest effect. (b) partial $\eta^2$ ANOVA effect size. Light blue indicates upper bound of bootstrapped 95% confidence interval. (c) shows Receiver-Operating-Characteristic (ROC) Curve for Logistic Regression classification. (d) shows distributional overlap, balanced accuracy (BACC) as well as sensitivity and specificity of the Logistic Regression classification.*

## Supplementary Figure S6. Predictive utility – RS connectivity - healthy versus depressive subjects

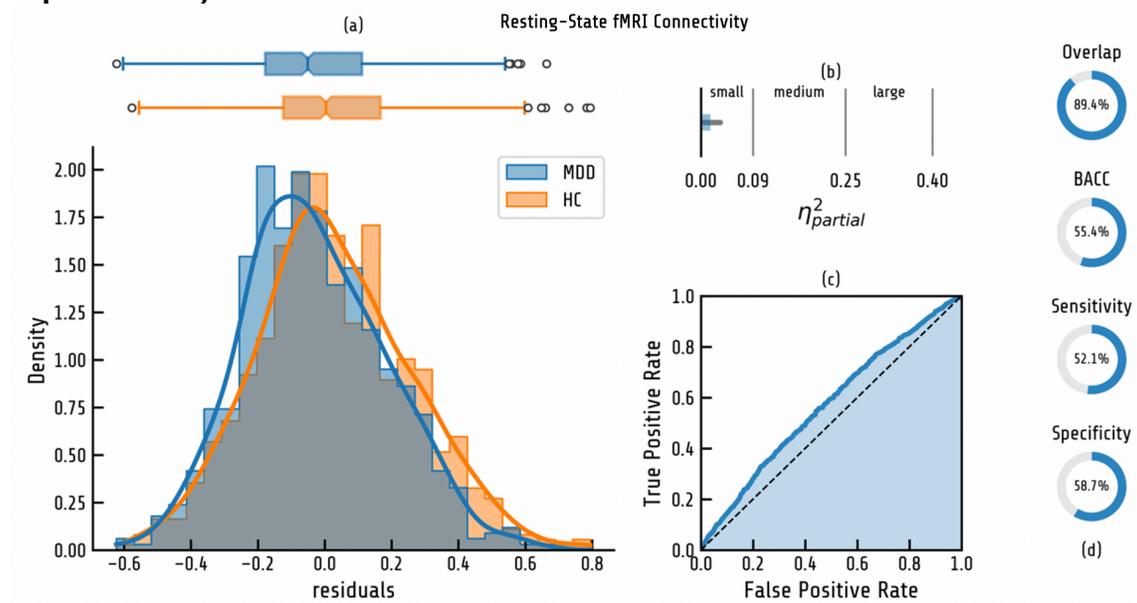

Figure S 6. Distributional overlap, effect size and classification performance for resting state (RS) fMRI connectivity data. (a) shows a histogram with Gaussian Kernel Density estimation as solid line and boxplot of the confound-corrected values of the connection displaying the largest effect. (b) partial $\eta^2$ ANOVA effect size. Light blue indicates upper bound of bootstrapped 95% confidence interval. (c) shows Receiver-Operating-Characteristic (ROC) Curve for Logistic Regression classification. (d) shows distributional overlap, balanced accuracy (BACC) as well as sensitivity and specificity of the Logistic Regression classification.

## Supplementary Figure S7. Predictive utility – RS graph - healthy versus depressive subjects

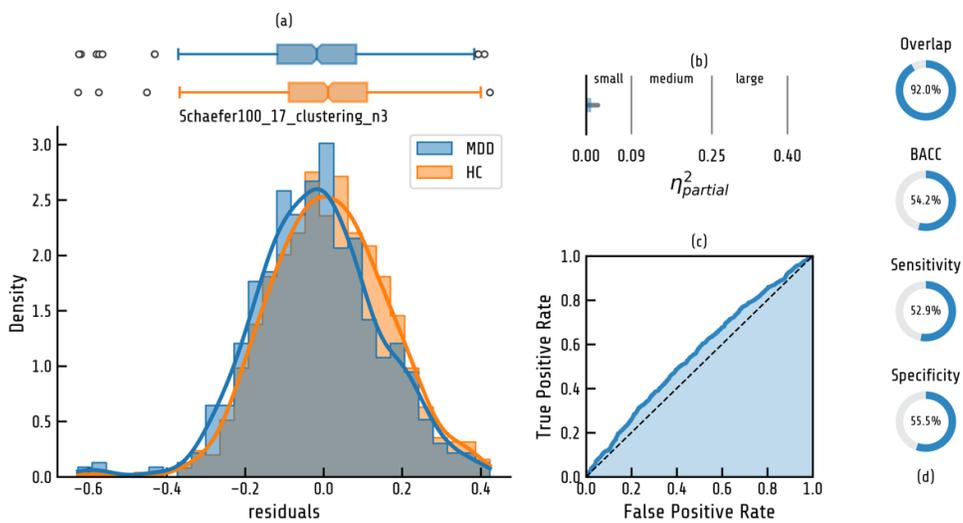

Figure S 7. Distributional overlap, effect size and classification performance for resting state (RS) fMRI network graph metrics. (a) shows a histogram with Gaussian Kernel Density estimation as solid line and boxplot of the confound-corrected values of the network graph parameter displaying the largest effect. (b) partial $\eta^2$ ANOVA effect size. Light blue indicates upper bound of bootstrapped 95% confidence interval. (c) shows Receiver-Operating-Characteristic

(ROC) Curve for Logistic Regression classification. (d) shows distributional overlap, balanced accuracy (BACC) as well as sensitivity and specificity of the Logistic Regression classification.

## Supplementary Figure S8. Effect size and classification accuracy - healthy versus acutely depressed subjects

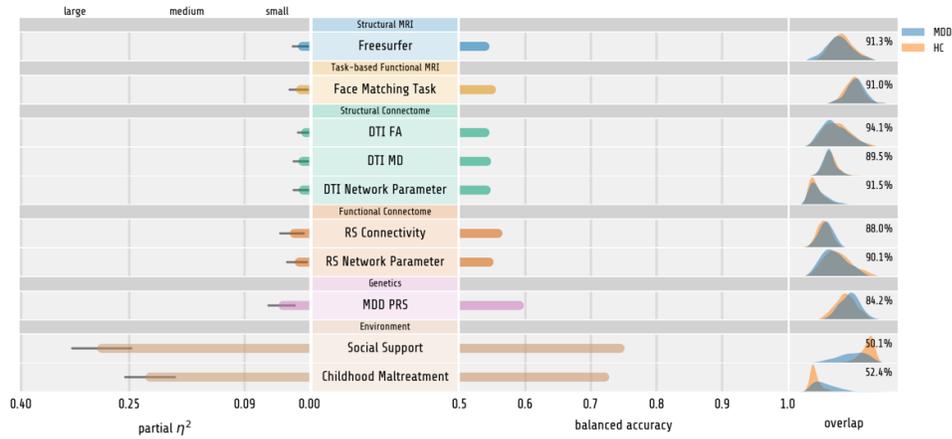

Figure S 8. Partial $\eta^2$ effect size of variables with highest F-value for all modalities. Error bars indicate upper bound for bootstrapped confidence intervals for partial $\eta^2$. Balanced classification accuracy for all modalities based on Logistic Regression of variables with lowest p-value. Kernel density estimation plots of deconfounded values including distributional overlap for healthy and depressive participants are plotted on the right side of the figure. HC = healthy controls, MDD = Major Depressive Disorder, DTI = Diffusion Tensor Imaging, FA = Fractional Anisotropy, MD = Mean Diffusivity, RS = Resting State functional MRI.

## Supplementary Figure S9. Effect size and classification accuracy - healthy versus chronically depressed subjects

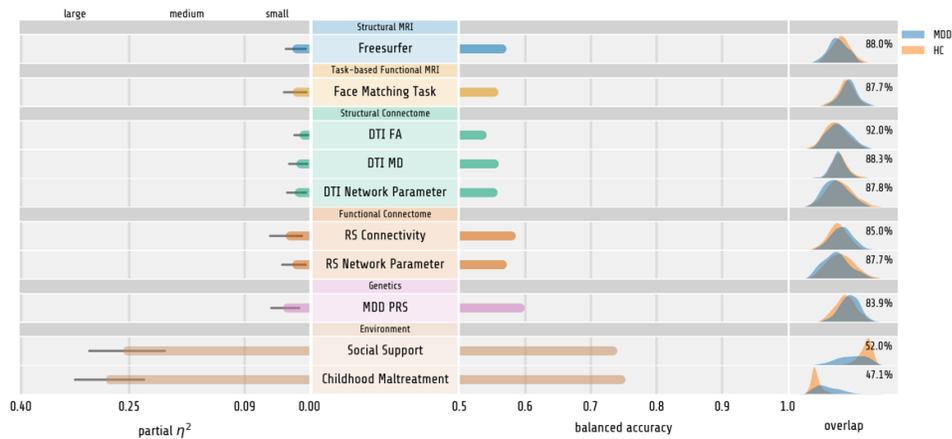

Figure S 9. Partial $\eta^2$ effect size of variables with highest F-value for all modalities. Error bars indicate upper bound for bootstrapped confidence intervals for partial $\eta^2$. Balanced classification accuracy for all modalities based on Logistic Regression of variables with lowest p-value. Kernel density estimation plots of deconfounded values including distributional overlap for healthy and depressive participants are plotted on the right side of the figure. HC = healthy controls, MDD = Major Depressive Disorder, DTI = Diffusion Tensor Imaging, FA = Fractional Anisotropy, MD = Mean Diffusivity, RS = Resting State functional MRI.


\end{document}

%% file: 01_Article_MainText.tex
\leadauthor{Winter}


\title{More Alike Than Different: Quantifying Deviations of Brain Structure and Function in Major Depressive Disorder across Neuroimaging Modalities}
\shorttitle{More Alike Than Different}

\author[1]{Nils R. Winter}
\author[1,2]{Ramona Leenings}
\author[1,2]{Jan Ernsting}
\author[1]{Kelvin Sarink}
\author[1]{Lukas Fisch}
\author[1]{Daniel Emden}
\author[1]{Julian Blanke}
\author[1]{Janik Goltermann}
\author[1]{Nils Opel}
\author[1]{Carlotta Barkhau}
\author[1,3]{Susanne Meinert}
\author[1]{Katharina Dohm}
\author[1]{Jonathan Repple}
\author[1]{Marco Mauritz}
\author[1]{Marius Gruber}
\author[1]{Elisabeth J. Leehr}
\author[1]{Dominik Grotegerd}
\author[1,4]{Ronny Redlich}
\author[5]{Andreas Jansen}
\author[5]{Igor Nenadic}
\author[6,7]{Markus Nöthen}
\author[7]{Andreas Forstner}
\author[8]{Marcella Rietschel}
\author[9]{Joachim Groß}
\author[10]{Jochen Bauer}
\author[10]{Walter Heindel}
\author[11]{Till Andlauer}
\author[12,13]{Simon Eickhoff}
\author[5]{Tilo Kircher}
\author[1*]{Udo Dannlowski}
\author[1*]{Tim Hahn}

\affil[1]{Institute for Translational Psychiatry, University of Münster, Münster, Germany}
\affil[2]{Faculty of Mathematics and Computer Science, University of Münster, Münster, Germany}
\affil[3]{Institute for Translational Neuroscience, University of Münster, Münster, Germany}
\affil[4]{Institute of Psychology, University of Halle, Halle, Germany }
\affil[5]{Department of Psychiatry and Psychotherapy, Phillips University Marburg, Marburg, Germany}
\affil[6]{Institute of Human Genetics, University of Bonn School of Medicine \& University Hospital Bonn, Bonn, Germany}
\affil[7]{Department of Genomics, Life \& Brain Research Center, University of Bonn, Bonn, Germany}
\affil[8]{Department of Genetic Epidemiology in Psychiatry, Central Institute of Mental Health, Medical Faculty Mannheim, University of Heidelberg, Mannheim, Germany}
\affil[9]{Institute for Biomagnetism and Biosignalanalysis, University of Münster, Münster, Germany}
\affil[10]{Institute of Clinical Radiology, University of Münster, Münster, Germany}
\affil[11]{Department of Translational Research in Psychiatry, Max Planck Institute of Psychiatry, Munich, Germany}
\affil[12]{Institute of Neuroscience and Medicine (INM-7), Research Centre Jülich, Jülich, Germany}
\affil[13]{Institute of Systems Neuroscience, Medical Faculty, Heinrich Heine University Düsseldorf, Düsseldorf, Germany}
\affil[*]{These authors contributed equally.}

\maketitle

\begin{abstract}
\underline{Introduction} - Identifying neurobiological differences between patients suffering from Major Depressive Disorder (MDD) and healthy individuals has been a mainstay of clinical neuroscience for decades. However, recent meta- and mega-analyses have raised concerns regarding the replicability and clinical relevance of brain alterations in depression.\newline 
\underline{Methods} - Here, we systematically investigate healthy controls and MDD patients across a comprehensive range of modalities including structural magnetic resonance imaging (MRI), diffusion tensor imaging, functional task-based and resting-state MRI under near-ideal conditions. To this end, we quantify the upper bounds of univariate effect sizes, predictive utility, and distributional dissimilarity in a fully harmonized cohort of N=1,809 participants. We compare the results to an MDD polygenic risk score (PRS) and environmental variables.\newline 
\underline{Results} - The upper bound of the effect sizes range from partial $\eta^2$ = 0.004 to 0.017, distributions overlap between 89\% and 95\%, with classification accuracies ranging between 54\% and 55\% across neuroimaging modalities. This pattern remains virtually unchanged when considering only acutely or chronically depressed patients. Differences are comparable to those found for PRS, but substantially smaller than for environmental variables.\newline 
\underline{Discussion} - We provide a large-scale, multimodal analysis of univariate biological differences between MDD patients and controls and show that even under near-ideal conditions and for maximum biological differences, deviations are remarkably small, and similarity dominates.  We sketch an agenda for a new focus of future research in biological psychiatry facilitating quantitative, theory-driven research, an emphasis on multivariate machine learning approaches, as well as the utilization of ecologically valid phenotyping.\newline 
\end {abstract}

\begin{keywords}
    Major Depressive Disorder | structural MRI | functional MRI | Polygenic Risk Scores | Theories of Depression | Biomarker
\end{keywords}

\begin{corrauthor}
\newline
Nils Winter\newline
Institute for Translational Psychiatry, University of Münster, Germany
\newline Albert-Schweitzer-Campus 1, D-48149 Münster
\newline Phone:	+49 (0)2 51 / 83 – 51847, Fax: +49 (0)2 51 / 83 – 56612
\newline E-Mail: nils.r.winter\at uni-muenster.de
\end{corrauthor}


\section*{Introduction}
\begin{figure*}[!t]
    \centering
    \includegraphics[width=1\linewidth]{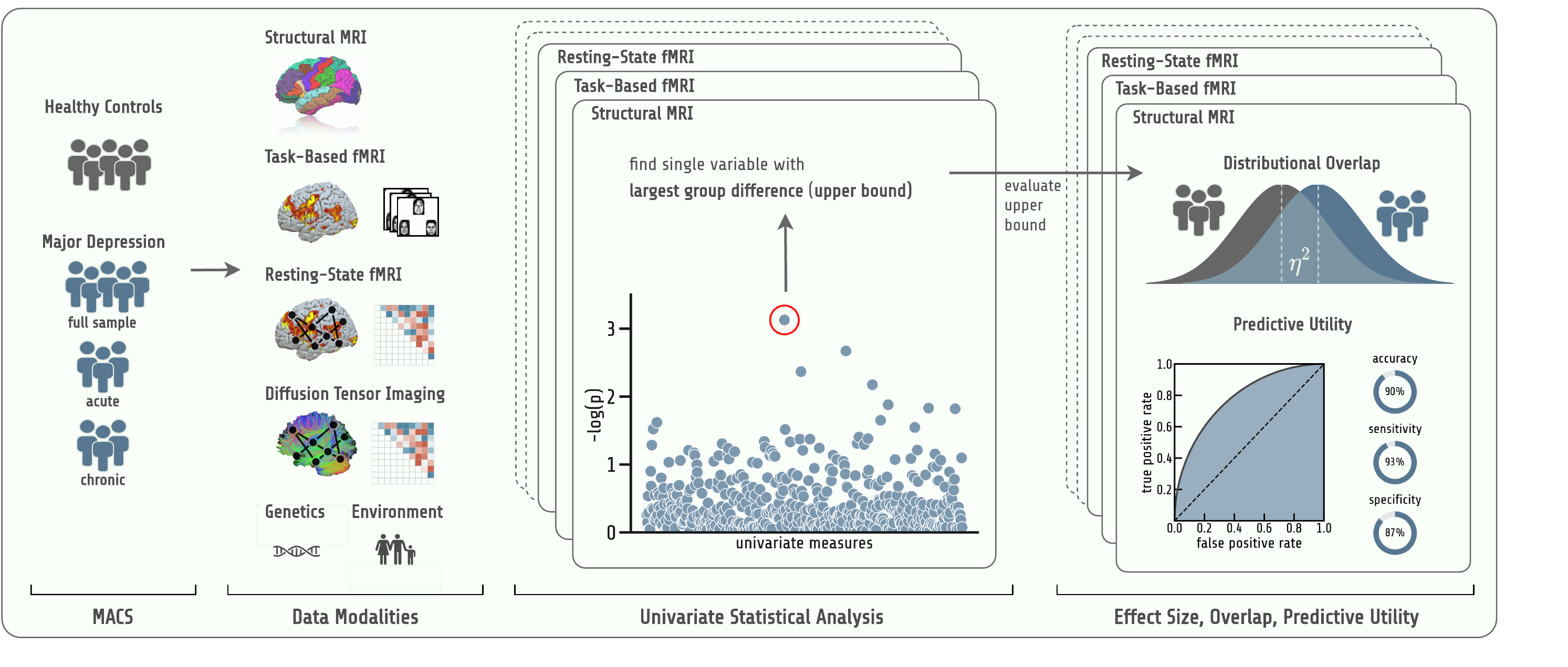}
    \caption{Schematic representation of the research design and analytical procedure. For all modalities, standard univariate models are calculated to find the single variable showing the largest difference between depressive and healthy individuals, representing an upper bound for univariate group differences. Then, effect size, distributional overlap and predictive utility is estimated for these peak variables of the different modalities. $\eta^2$= effect size of the statistical model. MACS = Marburg-Münster Affective Cohort Study. MRI = Magnetic Resonance Imaging. }
    \label{fig:MethodsDiagram}
 \end{figure*}

Major depressive disorder (MDD) is the single largest contributor to non-fatal health loss worldwide, annually affecting as many as 300 million people.\cite{Organization.2017} Globally, it resulted in a total estimate of over 50 million Years Lived with Disability (YLD), with MDD being one of the leading causes of YLD among all diseases.1 The incremental economic burden of adults is estimated to be over \$US320 billion in the US alone, including direct, suicide-related, and workplace costs; a notable increase by 37.9\% between 2010 and 2018.\cite{Greenberg.2021} \newline
Driven by the discovery of efficient psychopharmacological medication and the insight that many mental disorders have a strong genetic component, the second half of the twentieth century was dominated by biological psychiatry.\cite{Shorter.1998} With the emergence of cognitive neuroscience, neuroimaging, and neurogenetics, this paradigm evolved into a methodologically diverse systems medicine approach over the last two decades which views mental disorders as “relatively stable prototypical, dysfunctional patterns of experience and behavior that can be explained by dysfunctional neural systems at various levels”.\cite{Walter.2013} Correspondingly, identifying the neural and genetic basis of MDD to inform the improvement of treatments has been a mainstay of research in psychiatry for decades with over 1,500 neuroimaging studies listed on PubMed that investigate differences between healthy and depressed individuals (see Supplementary Methods for search strategy).
While large-scale projects and consortia accumulating neuroimaging data from tens of thousands of patients aim to consolidate and extend our understanding of mental disorders, there is growing concern regarding the replicability and prognostic utility of biomarkers derived from standard univariate analysis frameworks in psychiatry.\cite{Mueller.2016, Gray.2020, Schmaal.2016} Specifically investigating the neurobiological underpinnings of MDD, Müller et al. conducted a meta-analysis of 99 functional neuroimaging studies and found no spatial convergence of MDD effects vs. healthy subjects in paradigm-based fMRI studies.8 In the same vein, the meta-analysis by Gray et al. – based on 92 publications representing 2,928 patients with MDD – found no convergent neural alteration in depression compared to healthy individuals based on structural and functional MRI data.\cite{Mueller.2016} 
Investigating subcortical brain structures, a meta-analysis by the ENIGMA consortium showed a significant difference between 1,728 MDD patients and 7,199 controls from 15 samples worldwide.\cite{Schmaal.2016} This effect, however, was restricted to hippocampus volume and corresponds to a classification accuracy of merely 52.6\%; prompting Fried and Kievit to conclude that the result “leaves little hope to robustly distinguish between MDD and healthy participants based on univariate analyses of regional volumes”.\cite{Fried.2016h9} Furthermore, Malhi et al. comment that the “reiteration of non-speciﬁc volumetric changes does little to advance our knowledge of the illness”.\cite{Malhi.2016} In a later meta-analysis of cortical changes in depression, Schmaal et al. did find a larger number of regions that differ in cortical thickness between healthy and depressed individuals.\cite{Schmaal.2017} However, although these differences seem to be more widespread across the cortex and more regions reach significance, the associated effect sizes remain small and comparable to the magnitude of hippocampal changes. This general notion was equally apparent in white matter disturbances in MDD.\cite{Velzen.2020}
This lack of consistent findings and surprisingly small effects have been attributed to, first, methodological heterogeneity including varying experimental designs, varying inclusion and exclusion criteria, or meta-analytic approaches and, second, to the heterogeneity of the clinical population and its assessment including different severity, disease duration, or number of previous episodes.\cite{Gray.2020, Mueller.2016, Schmaal.2016yg8} \newline
Against this backdrop, the present study quantifies the magnitude and biomarker potential of univariate biological differences between depressive and healthy subjects across neuroimaging modalities in a fully harmonized study minimizing methodological and clinical heterogeneity. To this end, we draw on data of N=861 life-time or acute MDD patients and N=948 healthy controls (HC) from the bicentric Marburg-Münster Affective Cohort Study (MACS) comprising major neuroimaging modalities including cortical surface, cortical thickness and volume of brain regions based on structural MRI, task-based functional MRI, atlas-based connectivity and graph network parameters derived from resting-state functional MRI, fractional anisotropy (FA), mean diffusivity (MD) and graph network parameters based on Diffusion-Tensor-Imaging (DTI). For comparison, we also investigate an MDD PRS and environmental variables including self-reported childhood maltreatment and social support.\cite{Kircher.2019, Vogelbacher.2018}  Importantly, data from the MACS were fully harmonized, i.e., the same inclusion and exclusion criteria were applied, patients stemmed from the same country, thus sharing the same medical and socio-economic system, and the same psychometric assessments and interviews were conducted. Also, the same neuroimaging protocols – including a harmonization of scanner sequences – as well as quality control procedures were applied (for details, see \cite{Vogelbacher.2018}).\newline
In our analyses, we first assess statistical significance and corresponding effect sizes using established analysis standards for each modality (see Figure \ref{fig:MethodsDiagram}). For all modalities, we reported results of the single variable (i.e., score, voxel, graph metric, connectivity) displaying the largest difference between healthy and depressed individuals. Secondly, to gauge potential value as a biomarker of these variables showing the largest univariate difference, we quantified their predictive utility (i.e., accuracy, sensitivity, specificity) in every modality. Third, we illustrated the similarity between depressive and healthy participants with respect to the variable displaying the largest difference in every modality by calculating the overlapping coefficient (OVL), an intuitive measure of overlap between two populations.\cite{Inman.1989}  While focusing on the single largest variable is prone to overestimating the true difference, the approach provides a solid upper bound for the true deviation between healthy controls and MDD patients in the respective modality.\cite{Ferreira.2006}  
Explicitly investigating the substantial clinical heterogeneity often observed in MDD, we also conducted subgroup analyses including symptom severity, course of disease, gender, and scanner site. Finally, in order to benchmark the neuroimaging results, we compared all effects to the same set of metrics for genetic data (i.e., MDD PRS) as well as environmental factors in the same sample.

\section*{Methods}
\subsection*{Participants}
At the time of data analysis, 2036 healthy and depressed subjects participated in the cross-sectional Marburg-Münster Affective Cohort Study (MACS).\cite{Kircher.2019} Data were collected at two different sites (Marburg and Münster). After excluding subjects with any history of neurological (e.g., concussion, stroke, tumor, neuro-inflammatory diseases) and medical (e.g., cancer, chronic inflammatory or autoimmune diseases, heart diseases, diabetes mellitus, infections) conditions as well as non-Caucasian subjects, a final sample of 948 healthy and 861 depressed subjects (N=1809) were available for analysis. For the analysis of each individual data modality, all subjects for which data of the specific modality was available and passed quality checks were used. This resulted in a varying number of samples that could be used for the different modalities (see below). Patients with severe, moderate, mild or (partially) remitted MDD episodes were included irrespective of current treatment. A structured clinical interview for DSM-IV (SCID-I) was conducted with each participant in order to assess current and lifetime psychopathological diagnoses.\cite{Wittchen} Patients either fulfilled the DSM-IV criteria for an acute major depressive episode or had a lifetime history of a major depressive episode. 
For the secondary analyses including only acutely depressed patients, we removed fully remitted patients from the MDD sample (see Supplementary Methods for a definition of remission status). For the secondary analyses including only chronically depressed patients, we only used MDD subjects that had a history of at least two inpatient stays due to their depressive disorder.

\begin{figure*}[!t]
    \centering
    \includegraphics[width=1\linewidth]{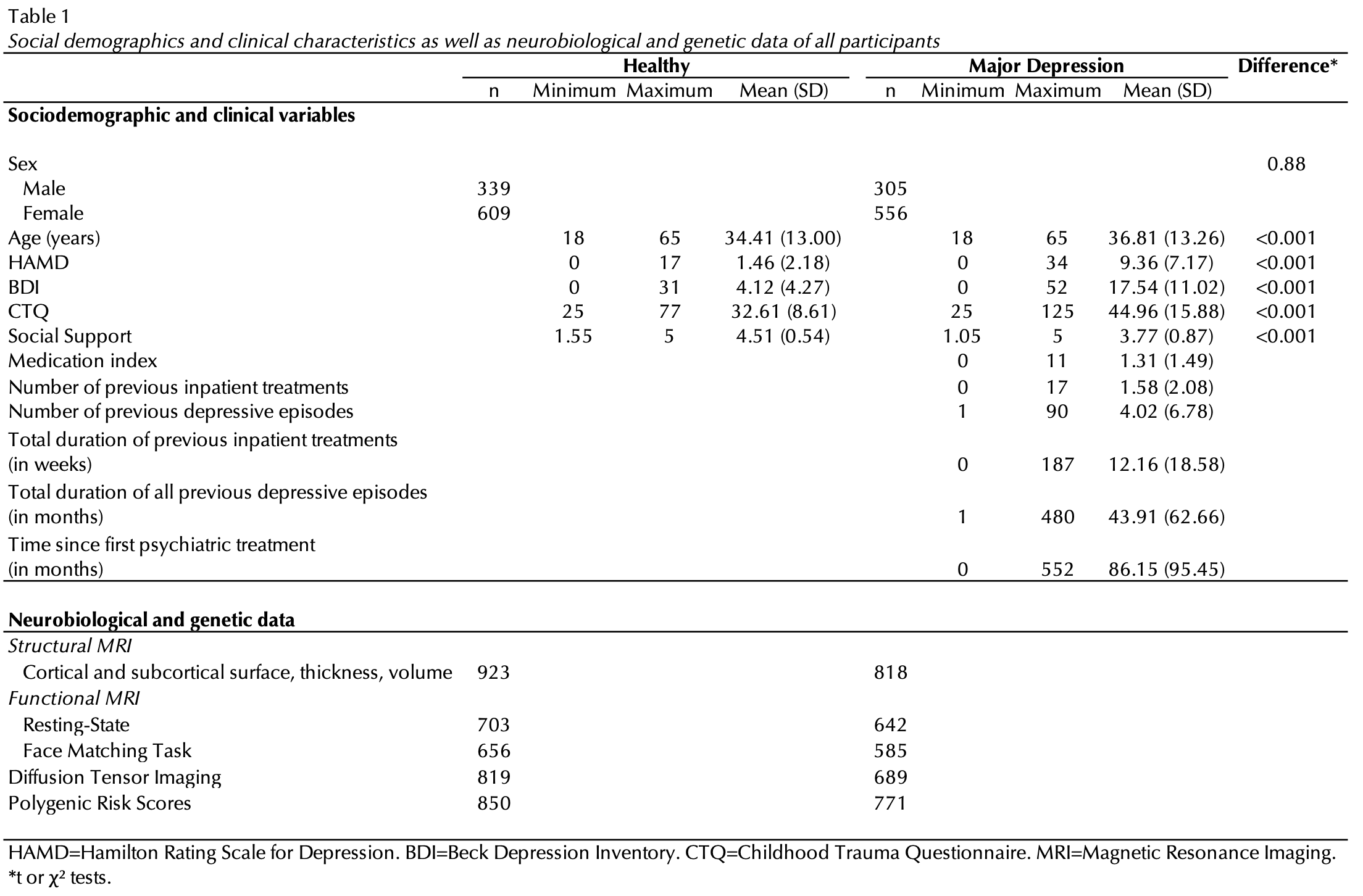}
    \label{fig:table_figure1}
\end{figure*}

\subsection*{Statistical analyses}
An Analysis of Variance (ANOVA) model predicting a single variable of interest was calculated for all variables of the different modalities with age, gender, and scanning site as a minimum set of covariates and a HC versus MDD factor. 
As the MRI body-coil was changed mid recruitment in Marburg, scanning site was dummy coded to represent three categories (pre and post body-coil change Marburg and Münster) and used in the Hariri task analysis.
For PRS analyses, the first three Multi-Dimensional Scaling (MDS) components of the genetic relationship matrix were added as covariates to correct for population stratification (for details, see \cite{Pelin.2021}). 
General linear models were calculated in Python using the statsmodels package.\cite{Seabold} Voxel-wise whole brain analyses were done using SPM12.\cite{Penny.2011}
For all analyses except the ones based on voxel-wise data (Hariri task fMRI), we controlled for multiple comparisons by calculating the false discovery rate with a false positive rate of 0.05 using the Benjamini-Yekutieli procedure.\cite{Benjamini.2001} For voxel-based data, significance thresholds for multiple testing were obtained at the voxel-level using non-parametric t tests as implemented in the SPM Threshold-Free Cluster Enhancement (TFCE) toolbox.\cite{Gaser.2019} An FWE-corrected threshold of 0.05 was used to calculate corrected p-values. We decided against a cluster correction as implemented in the TFCE toolbox since the peak voxel of a significant cluster must not necessarily show the strongest effect among all voxels in the brain. This is, of course, usually a desirable characteristic of the TFCE correction. However, as we intended to find the upper bound, i.e., the largest difference between healthy and depressive subjects, we used a correction method on a voxel-level. For each modality, the variable showing the strongest effect (largest F-value and thus lowest p-value) was selected and $\eta^2_{partial}$ was calculated as measure of effect size for the group factor (HC versus MDD). Bootstrap confidence intervals were calculated using the Bias-corrected and accelerated (BCa) bootstrap method including group stratification.\cite{Diciccio.1988} 
To quantify the predictive potential of the variables showing the largest group effect, a logistic regression was fitted on the deconfounded residuals of the linear models. In other words, the covariates used in the respective modality analyses were regressed out of the data using the same linear models as described earlier, only removing the group factor (diagnosis) from the model first. The predictions of the logistic regression were then used to compute balanced accuracy, sensitivity and specificity. Additionally, the probabilities which can be obtained from the logistic regression model were used to plot a Receiver Operating Characteristic Curve (ROC). All code implementing the statistical analyses and figures is publicly available under \url{https://github.com/wwu-mmll/more-alike-than-different-paper2021}.

\subsection*{Assessment of childhood maltreatment and social support}
The well-established childhood trauma questionnaire (CTQ) was used to assess childhood maltreatment in patients and controls.\cite{Bernstein.1994} For all analyses including childhood maltreatment, we used a sum score that can be computed across the five maltreatment subscales. For 12 subjects no CTQ sum score was available, resulting in a sample of 1797 subjects that could be used for the CTQ analysis.
Perceived social support was measured using the Social Support Questionnaire (F-SozU), an established German self-report instrument.\cite{Fydrich.2009} The questionnaire consists of three subscales measuring the subject’s perceived emotional support, instrumental support, and social integration. A sum score aggregating all three scales was used in all analyses. For 10 subjects no social support sum score was available, resulting in a sample of 1799 subjects that could be used for the social support analysis.

\subsection*{Polygenic risk score for depression}
Genotyping was conducted using the PsychArray BeadChip, followed by quality control and imputation, as described previously.\cite{Andlauer.2016, Meller.2019} In brief, QC and population substructure analyses were performed in PLINK v1.9.\cite{Chang.2015} Imputed genetic data were available for n=1689 individuals. Related subjects were identified using PLINK and one individual of each related pair was excluded on the analysis level. This procedure was done separately for every analysis to guarantee that only a minimal number of related samples were excluded for every MDD, gender, and site subgroup analysis. A final sample of 1621 was available for the HC versus MDD analysis. Sample sizes for all other analyses are listed in Table \ref{fig:table_figure1}. For the calculation of the PRS, single nucleotide polymorphism weights were estimated using the PRS-CS method with default parameters (for details, see Supplementary Methods).\cite{Andlauer.2020, Ge.2019}  A PRS-CS for Major Depression was calculated using summary statistics from a recent GWAS.\cite{Howard.2019} The global shrinkage parameter $\phi$ was automatically determined (PRS-CS-auto; $\phi=1.30*10^{-4}$). To control for population substructure, three MDS components were added as covariates to all linear models containing genetic data. 

\subsection*{Magnetic Resonance Imaging}
Magnet resonance imaging (MRI) data for all brain-based modalities were acquired using two 3T whole body MRI scanner (Marburg: Tim Trio, Siemens, Erlangen, Germany; Münster: Prisma, Siemens, Erlangen, Germany). Due to a change of the scanner body-coil in Marburg during recruitment, three different scanner conditions were used as covariate in all MRI-based analyses (Marburg pre body-coil change, Marburg post body-coil change and Münster). Imaging protocols were harmonized across scanner sites to the extent permitted by each platform. Pulse sequence parameters as well as quality assurance protocols have been described in detail previously.\cite{Vogelbacher.2018} Additional information on MRI scanning parameters is provided in the Supplementary Methods.

\subsection*{Cortical and subcortical segmentation of T1-weighted MRI}
Automated segmentation was conducted using the cortical and subcortical parcellation stream of Freesurfer (Version 5.3) based on the Desikan-Killiany atlas.\cite{Fischl.2012} In total, measures for 68 cortical regions (34 on each hemisphere), 14 subcortical regions (7 on each hemisphere), 4 ventricles (2 on each hemisphere), as well as total intracranial volume (ICV) were extracted for each participant. Additionally, global measures of cortical and subcortical surface, thickness and volume were calculated per and across hemisphere (for a full list of used Freesurfer parameters, see Supplementary Methods). This resulted in a total number of 166 parameters that were used in the statistical analyses. Default parameters were used for the segmentation (\url{https://surfer.nmr.mgh.harvard.edu/}) and segmentation quality was reviewed visually as well as based on statistical outlier analysis following standardized protocols by the ENIGMA consortium.\cite{Consortiumptj} After excluding subjects with poor segmentation quality, a final sample of 1741 subjects were available for the Freesurfer analysis. 

\subsection*{Resting-state fMRI}
Two-hundred-thirty-seven interleaved and ascending measurements were acquired for resting-state fMRI. Participants were asked to keep their eyes closed until the end of the resting state session. Resting-state preprocessing was done using the CONN (v18b) MATLAB toolbox and the default volume-based MNI preprocessing pipeline.\cite{Whitfield-Gabrieli.2012} First, a series of preprocessing steps was applied to the functional and structural images that included a functional realignment and unwarp, a slice-timing correction, an ART-based outlier identification, a direct segmentation and normalization of the functional and structural images, as well as a final functional smoothing with an 8mm FWHM kernel. Second, CONN’s denoising step with default parameters was applied to regress out potential noise artefacts in the functional data based on an anatomical component-based noise correction procedure (aCompCor). Controlling for noise artefacts is done by regressing out noise components from cerebral white matter and cerebrospinal areas (5 PCA components), estimated subject-motion parameters (12 parameters including 6 motion parameters and their associated first-order derivatives), as well as identified outlier scans or scrubbing. Finally, temporal band-pass filtering was applied to remove low frequencies under 0.008 Hz and high frequencies above 0.09 Hz.
Connectivity matrices for each subject were created by computing the bivariate Pearson’s correlation coefficient between the time-series of every region of the 17 networks Schaefer atlas with 100 parcels\cite{Schaefer.2017} Resting-state and T1 data was available for 1374 subjects. 12 subjects were excluded because of an acute medication with tranquilizers such as benzodiazepine or Z-drugs which are known to influence functional MRI data. 17 subjects were excluded due to strong motion artifacts that resulted in less than 5 minutes of valid resting-state image time points after scrubbing. Data of a final sample of 1345 subjects was available for resting-state analyses.

\begin{figure*}[!t]
    \centering
    \includegraphics[width=1\linewidth]{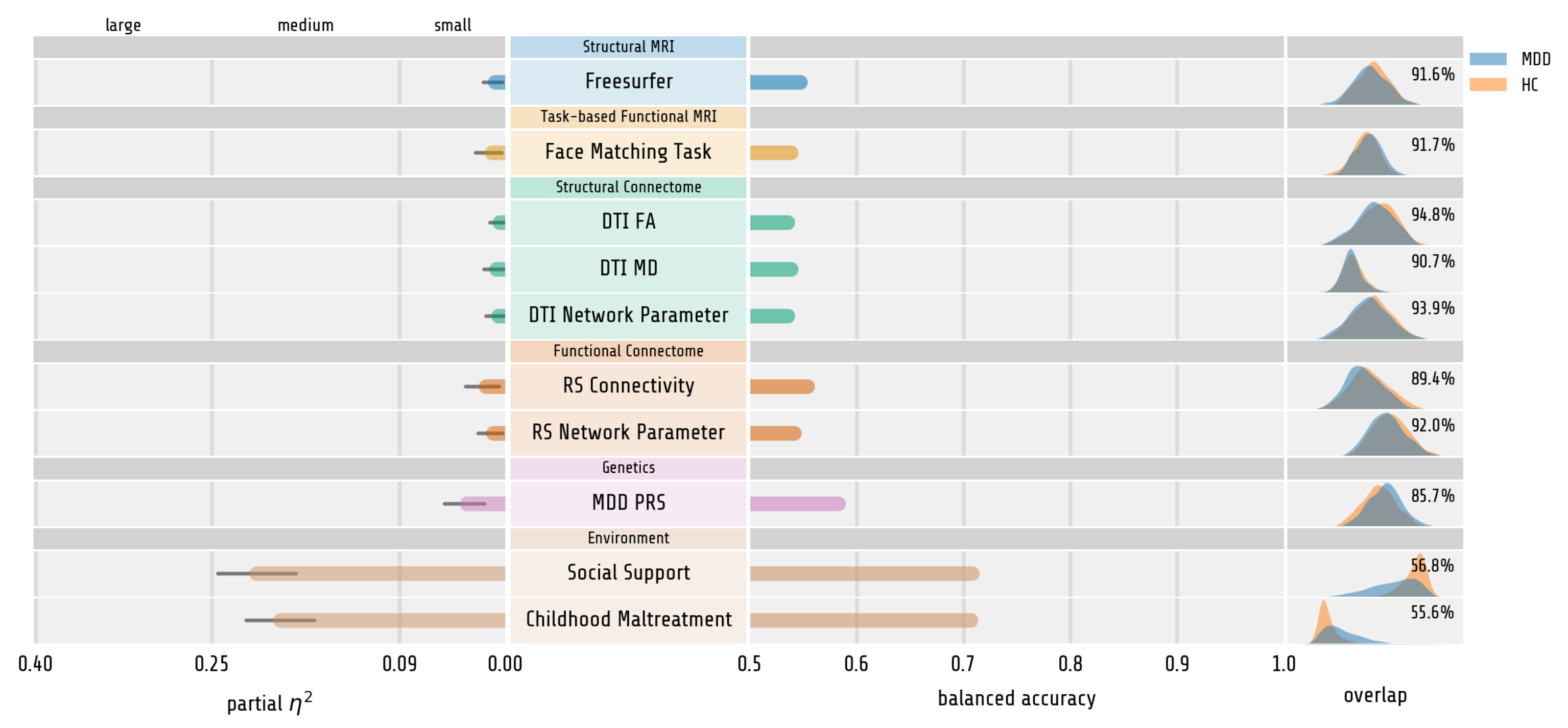}
    \caption{Left: $\eta^2_{partial}$ effect size of single variables displaying the overall largest effect in each modality. Error bars indicate upper and lower bound for bootstrapped confidence intervals for $\eta^2_{partial}$. Right: Balanced classification accuracy for all modalities based on Logistic Regression of single variables displaying the largest effect. Kernel density estimation plots of deconfounded values including distributional overlap for healthy and depressive participants are plotted on the right side of the figure. HC = healthy controls, MDD = Major Depressive Disorder, DTI = Diffusion Tensor Imaging, FA = Fractional Anisotropy, MD = Mean Diffusivity, RS = Resting State functional MRI. }
    \label{fig:accuracy}
\end{figure*}

\subsection*{Task-based fMRI}
Functional MRI data from a well-established emotional face matching paradigm was used.\cite{Hariri.2002} The experimental setup and preprocessing have been described previously.\cite{Kessler.2020} In short, subjects viewed images of fearful or angry faces in the experimental and geometric shapes in the control condition. In each trial, a target image was presented at the top while two further images were presented at the bottom left and right, whereby one of these images was identical to the target image. The subjects were instructed to indicate whether the left or right image was identical to the top image by pressing a corresponding button. \newline
Image acquisition details corresponded to the previously described resting-state scanning parameters. To avoid motion artifacts, subjects were excluded from the final sample if their overall movement exceeded 2mm. Additionally, a visual quality check has been conducted to exclude subjects with visually detectable artifacts. As described above, subjects under acute medication with tranquilizers have been excluded from the analyses. At the individual subject level, fMRI responses for both conditions (faces, shapes) were modeled in a block design using the canonical hemodynamic response function implemented in SPM8 convolved with a vector of onset times for the different stimulus blocks. High-pass filtering was applied with a cut-off frequency of 1/128 Hz to attenuate low-frequency components. Contrast images were created by contrasting beta images of the ‘faces’ against the ‘shapes’ condition. A sample of 1241 subjects were available for task-based fMRI analyses.

\subsection*{Diffusion Tensor Imaging}
Data were acquired using a GRAPPA acceleration factor of 2. Fifty-six axial slices with no gap were measured with an isotropic voxel size of 2.5×2.5×2.5mm³ (TE=90ms, TR=7300ms). Five non-DW images ($b=0 s/mm^2$) and 2×30 DW images with a b-value of $1000 s/mm^2$ were acquired. For additional image acquisition and preprocessing details, see Supplementary Methods. 
DTI image preprocessing was done as described in \cite{Repple.2020}. For every subject, the network information was stored in a structural connectivity matrix, with rows and columns reflecting 114 cortical brain regions of the Lausanne parcellation, a subdivision of the FreeSurfer’s Desikan-Killiany atlas.\cite{Hagmann.2008, Cammoun.2012iq} Matrix entries represent the weights of the graph edges. Network edges were weighted according to fractional anisotropy (FA), mean diffusivity (MD), and number of streamlines (NOS). For all FA- and MD-based connectivity analyses, only edges with non-zero values for at least 95\% of subjects were analyzed. This resulted in a varying number of subjects that were included in the statistical modeling of group differences in edge values. Data of 1508 subjects were available for all DTI-based analyses.

\subsection*{Graph network parameters}
For the DTI connectivity matrix, a binary adjacency matrix was calculated on the basis of number of streamlines. All edges with less than three number of streamlines were set to 0, all other edges were set to 1. For the resting-state fMRI connectivity matrix, a binary adjacency matrix was calculated by setting the top 15 percent of connections (highest correlation coefficient) to 1, all other edges were set to 0. The adjacency matrices were then used to calculate a number of representative graph parameters using PHOTONAI Graph (\url{https://github.com/wwu-mmll/photonai_graph}), which itself calls function from the Python package networkx.\cite{Hagberg} Used graph metrics were defined as follows (for an introduction of graph metrics for brain connectivity, see \cite{Rubinov.2010}): Global efficiency was defined as the average inverse shortest path length between all node pairs. Local efficiency was defined as the global efficiency on node neighborhoods. Clustering coefficient was computed as the average likelihood that the neighbors of a node are also mutually connected. Degree centrality was defined as the number of nodes connected to the node of interest. Betweenness centrality was defined as the fraction of shortest paths of the network that pass through the node of interest. Degree assortativity was defined as the similarity of connections in the graph with respect to the node degree. Clustering coefficient, degree centrality, and betweenness centrality were calculated per node and additionally averaged across nodes, resulting in a total of 348 network parameters.

\section*{Results}
 \begin{figure*}[!t]
    \centering
    \includegraphics[width=0.8\linewidth]{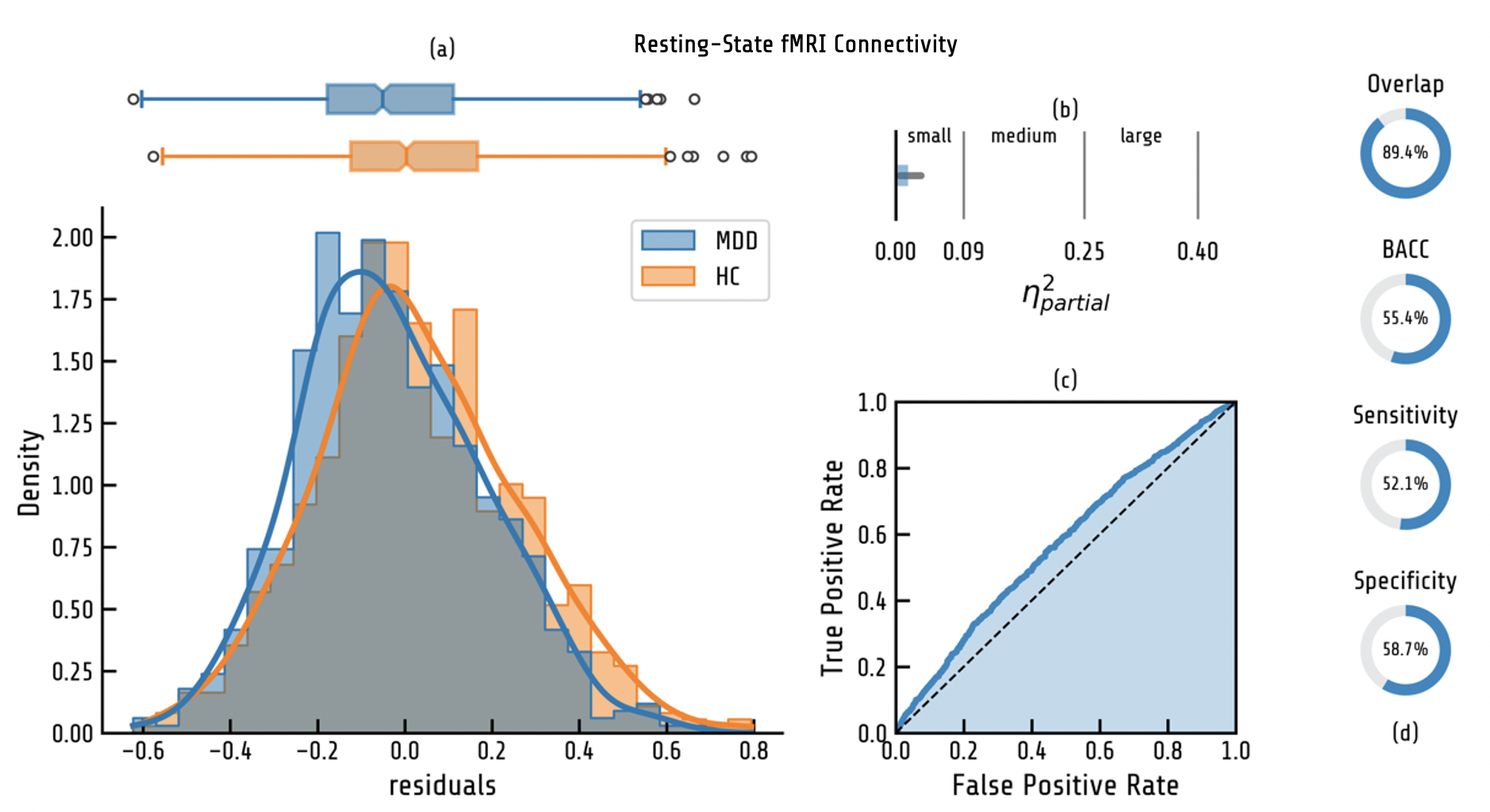}
    \caption{Distributional overlap, effect size and classification performance for the single variable displaying the largest effect across all neuroimaging modalities. (a) shows a histogram with Gaussian Kernel Density estimation as solid line and boxplot of the confound-corrected values of the resting-state ROI-to-ROI connectivity displaying the largest effect. (b) $\eta^2_{partial}$ ANOVA effect size for the strongest resting-state connectivity. Light blue indicates upper bound of bootstrapped 95\% confidence interval. (c) shows Receiver-Operating-Characteristic (ROC) Curve for Logistic Regression classification based on the single variable displaying the largest effect. (d) shows distributional overlap, balanced accuracy (BACC) as well as sensitivity and specificity of the Logistic Regression classification.  }
    \label{fig:resting-state}
 \end{figure*}

\subsection*{Effect sizes, distributional overlap and classification performance for HC versus MDD}
For the single variables displaying the largest difference between HC and MDD, ANOVA effect sizes were small in all neuroimaging modalities. They ranged from $\eta^2_{partial}=0.004$ for the largest effect in DTI data to $\eta^2_{partial}=0.017$ for the largest effect in resting-state connectivity (see Figure \ref{fig:accuracy}). For structural MRI, the greatest difference between healthy and depressed subjects could be observed in the total cortical volume of the right hemisphere (Freesurfer, $p_{uncorr} = 1.16*10^{-4}$, $p_{corr} = 0.063$, $F_{(1, 1735)} = 14.92$, $\eta^2_{partial} = 0.009$, see Supplementary Figure S1). For task-based functional MRI, the greatest difference in brain activation during a face matching task between healthy and depressed subjects were observed in a voxel within the left superior frontal region ($p_{uncorr} = 1.67*10^{-4}$, $p_{corr} = 0.398$, $F_{(1, 1235)} = 14.3$, $\eta^2_{partial} = 0.011$, see Supplementary Figure S2). For DTI, the greatest difference between healthy and depressed subjects was found for FA between the right pars triangularis and right rostral middle frontal region ($p_{uncorr} = 0.010$, $p_{corr} = 1$, $F_{(1, 1496)} = 6.71$, $\eta^2_{partial} = 0.004$, see Supplementary Figure S3), MD ($p_{uncorr} = 0.001$, $p_{corr} = 1$, $F_{(1, 1494)} = 11.20$, $\eta^2_{partial} = 0.007$, see Supplementary Figure S4) and the average degree centrality network parameter ($p_{uncorr} = 0.003$, $p_{corr} = 1$, $F_{(1, 1502)} = 8.93$, $\eta^2_{partial} = 0.006$, see Supplementary Figure S5). For resting-state functional MRI, the greatest difference between healthy and depressed subjects was measured  for the connectivity between a region of the right peripheral visual network and a region of the somato-motor network A ($p_{uncorr} = 2.24*10^{-6}$, $p_{corr} = 0.051$, $F_{(1, 1339)} = 22.57$, $\eta^2_{partial} = 0.017$, see Supplementary Figure S6) and the clustering coefficient of region 3 (central visual network) of the Schaefer 100 atlas ($p_{uncorr} = 0.001$, $p_{corr} = 1$, $F_{(1, 1339)} = 11.20$, $\eta^2_{partial} = 0.007$, see Supplementary Figure S7). \newline
In comparison to the neuroimaging data, healthy and depressed subjects differed significantly in the PRS for major depression ($p = 2.92*10^{-13}$, $F_{(1, 1613)} = 20.56$, $\eta^2_{partial} = 0.032$) and in social support ($p = 8.51*10^{-95}$, $F_{(1, 1792)} = 481.93$, $\eta^2_{partial} = 0.211$) and childhood maltreatment ($p = 8.08*10^{-85}$, $F_{(1, 1794)} = 425.59$, $\eta^2_{partial} = 0.192$).
Distributions of the variables displaying the largest difference between HC and MDD overlap between 89.4\% and 94.8\% across all neuroimaging modalities (see Figure \ref{fig:accuracy}). Even under ideal statistical conditions, this corresponds to classification accuracies between 53.5\% and 55.4\%. Lowest classification accuracy was found for the largest effect in DTI FA while the largest effect in resting-state connectivity displayed the highest classification accuracy (see Figure \ref{fig:resting-state}). In comparison, MDD PRS was found to have an overlap of 85.7\%, corresponding to a classification accuracy of 58.3\%. In contrast, environmental variables showed an overlap of 55.6\% and 56.8\%, corresponding to a classification accuracy of 70.7\% and 70.8\%. \newline
To further analyze the effect of heterogeneity due to research site or gender, we repeated all analyses for the two study sites Marburg and Münster as well as males and females separately. Although methodological and biological homogeneity were expected to increase within the respective subsamples, results did not fundamentally change (Supplementary Table S1-2).

 \begin{figure*}[!t]
    \centering
    \includegraphics[width=1\linewidth]{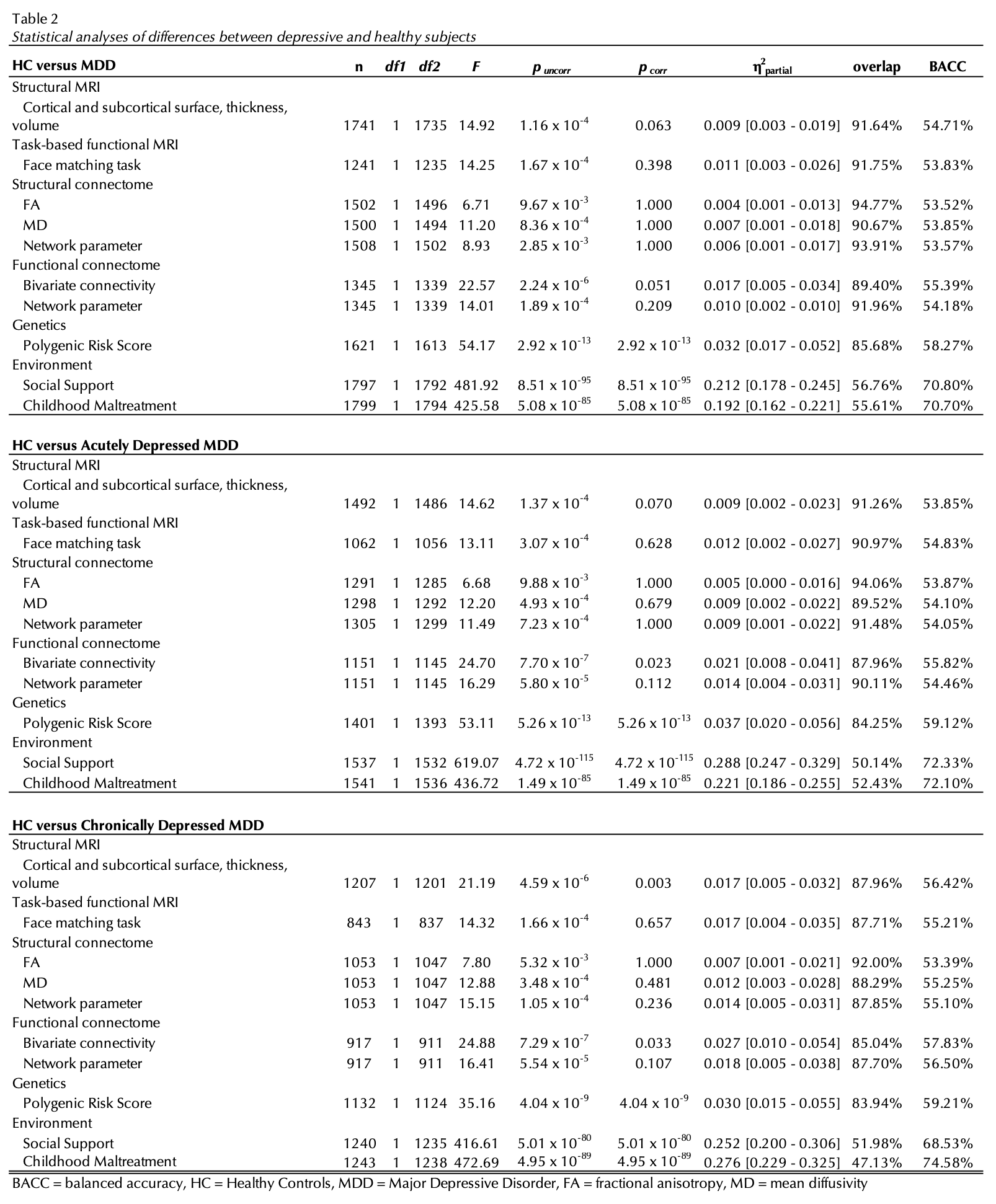}
    \label{fig:table2}
 \end{figure*}
 
\subsection*{Analysis of acutely and chronically depressed subgroups}
For the variables displaying the largest group difference in each modality, distributions of healthy and acutely depressed individuals overlapped between 87.9\% and 94.1\% for all neuroimaging modalities (see Table \ref{fig:table2}). Classification accuracies ranged between 53.9\% and 57.7\% for those variables displaying the maximum effect. Largest effect size was found within resting-state connectivity ($\eta^2_{partial} = 0.021$). To compare, overlap between healthy and acutely depressed individuals was 84.3\% for PRS, corresponding to a classification accuracy of 59.1\% and an effect size of $\eta^2_{partial} = 0.037$. In contrast, overlap for social support and childhood maltreatment was 50.1\% and 52.4\%, which corresponds to a classification accuracy of 72.3\% and 72.1\%. Effect sizes were moderate to large with $\eta^2_{partial} = 0.288$ for social support and $\eta^2_{partial} = 0.221$ for childhood maltreatment.
Comparably, distributions of maximum difference variables for healthy and chronically depressed individuals overlapped between 85.0\% and 92.0\% for all neuroimaging modalities (see Supplementary Table S1). Classification accuracies ranged between 53.4\% and 59.2\% for those variables displaying the maximum effect. Largest effect size was found within resting-state connectivity ($\eta^2_{partial} = 0.027$). To compare, overlap between healthy and chronically depressed individuals is 83.9\% for PRS, corresponding to a classification accuracy of 59.2\%. In contrast, overlap for social support and childhood maltreatment was 52.0\% and 47.1\%, which corresponds to a classification accuracy of 68.5\% and 74.6\%. Effect sizes were large with $\eta^2_{partial} = 0.252$ for social support and $\eta^2_{partial} = 0.276$ for childhood maltreatment.

\section*{Discussion}
In this study, we show that healthy and depressed individuals are strikingly similar with regard to univariate neurobiological and genetic measures. Even when considering the upper bound of the deviation in each modality, none could be considered informative from a personalized psychiatry perspective with both groups being nearly indistinguishable on a single-subject level. This is true despite near-ideal harmonization of study protocols, quality control, neuroimaging data acquisition, and clinical assessment, employing standard processing and analysis pipelines frequently used in the scientific community. Irrespective of their statistical significance – which was evident at least on uncorrected levels – even the upper bound of the effect size was small for all neuroimaging modalities. Overall, no modality explained more than ~2\% of the variance between healthy and depressive subjects. 
Our result for structural MRI data is in-line with Schmaal et al. who reported an explained variance of about 1\% for their largest effect of hippocampal volume reduction ($d=0.21$, $\eta^2 = 0.011$, see \cite{cohen}, page 22 for a transformation from $d$ to $\eta^2$). Importantly, however, we extend this finding to a comprehensive set of neuroimaging modalities and show that results are similar also for task-based functional MRI, atlas-based connectivity and graph network parameters derived from resting-state functional MRI, fractional anisotropy, mean diffusivity and graph network parameters based on DTI. Crucially, we show that the observed low effect sizes cannot be explained by a lack of harmonization of studies as previously suggested.\cite{Schmaal.2016yg8} Likewise, extensive subgroup analyses reveal that clinical heterogeneity is also not driving results (see below).\newline
Connecting these results from neuroimaging and psychiatric genetics, we show that MDD PRS only provided marginally higher effect size, explaining up to 5\% of the between-group variance. These results are in stark contrast to self-reported environmental factors such as childhood maltreatment or perceived social support which explain 6 to 48 times more interindividual variation compared to neuroimaging and genetic data. Likewise, the distributional similarity between healthy and depressive subjects even in the single variables displaying the largest difference is substantial. Converting these univariate differences to classification accuracy as a metric of predictive potential in a personalized psychiatry framework, they peak at about 59\% and do not exceed 55\% for most modalities. As this does not include an independent replication as required in predictive analyses, the prognostic and clinical value of these small effects are overly optimistic and - again – provide an upper bound of predictive utility. Importantly, secondary analyses on more clinically and methodologically homogeneous subgroups of acutely or chronically depressed, female or male individuals or individuals from only one scanning site did not change this pattern of results. In contrast to previous reports, our study leaves little room to attribute the lack of substantial differences between HC and MDD to small sample size or heterogeneity in study protocols and assessments. \newline
If the informational as well as the predictive value of cross-sectional, univariate group differences - as we show in this study - is negligible, we must first explain why we see such a similarity between neurobiological measurements of depressive and healthy subjects despite a substantial behavioral difference. Secondly, we need to derive ways of advancing the development of a theory of depression, changing the clinical practice and improving the well-being of patients.
\subsection*{How to explain the discrepancy between epidemiological reality of patients and a lack of neurobiological deviation?}
Contrasting our empirical findings with the epidemiological reality of patients’ substantial suffering, we will tackle the question of how to explain the lack of neurobiological deviation from a methodological and phenotypical view. In principle, it is possible that clinical neuroscience might simply be measuring properties of the brain which are irrelevant to MDD. Given the consistently small effects across all investigated modalities including brain structure, function and genetics, our results would suggest to direct research efforts towards brain measurements that are temporally and spatially more finely grained and could thus provide more clinically relevant information. More sophisticated neuroimaging techniques such as graph-theory-based connectome analyses, improved fMRI data acquisition, or high-field structural imaging might enhance the potential of neuroimaging data for providing univariate disease biomarkers.\cite{Repple.2020} Regarding fMRI, increasing evidence points to low reliability values particularly of task-based fMRI, and efforts have been suggested to address these issues.\cite{Elliott.2020}  While structural MRI does not encounter such reliability issues, standard volumetric analysis pipelines such as Freesurfer might not capture variance with direct relevance for disease-related mechanisms.\cite{Schmaal.2016, Schmaal.2017} Due to their high level of standardization, these structural data pipelines have, however, dominated large consortia such as ENIGMA lately.\cite{Elliott.2020} While more attainable, EEG and MEG, e.g., studies must build platforms and consortia to obtain similar sample-sizes as current large-scale structural MRI studies. \newline
Even if current neuroimaging data does contain clinically relevant information, standard univariate analysis approaches might not be able to adequately model the complexity of the depressive phenotype and underlying biological causalities. To reach a higher level of personalization in psychiatry, multivariate methods with their clear focus on predictive and clinical utility but also their ability to model complex relationships should become an even stronger part of the neuroimaging tool kit. Another advantage of these methods lies in their ability to integrate multiple data modalities, which could also increase the amount of relevant information within our models of depression. Complementary to the possibility of inadequate neurobiological measurements is the possibility that we might be considering ill-defined phenotypes. This notion is supported by the fact that the classification of disorders in psychiatry – be it in DSM or ICD – is geared towards reliability of diagnoses, and not neurobiological consistency.\cite{Cuthbert.2013} While numerous attempts such as the Research domain criteria (RDoC) have been suggested to assess phenotype characteristics in a manner as to make them more accessible to neuroscientific investigation, none have yielded substantial progress so far.\cite{Insel.2010}  The discussions around biomarkers preceding the publication of the DSM 5 strongly highlight this point.\cite{Scull.2021} Along the same lines, many have argued that depression is not a consistent syndrome with a fixed set of symptoms identical for all patients. In an investigation of patient’s symptoms in the STAR*D study, Fried and Nesse identified over 1000 unique symptoms in about 3,700 depressed patients, irrespective of depression severity.\cite{Fried.2015f8h} With these symptoms potentially differing from each other with respect to their underlying biology, severity, or impact on functioning, the common notion of aggregating across these diverse symptom profiles and focusing on MDD as homogeneous phenomenon has likely hampered the development of clinically useful biomarkers of depression.\cite{Fried.2015} In short, trying to associate a single biological variable with a complex disease category such as MDD, one can only find the variance that is shared across MDD patients. As only about 2\% of the patients in the STAR*D study belong to the most common symptom profile, this shared variance and subsequently the explained variance in a statistical model is likely to be small, which corresponds to and might explain the small effects we find.\cite{Fried.2015f8h} Still, depressive core symptoms  shared across patients clearly point to the existence of at least some common dysfunctions that also need to have a neurobiological basis we should be able to find. Thus, investigating the neurobiological basis of individual symptoms or dysfunctions is a promising research direction that receives increased attention.\cite{Fried.2015, Fried.2015t85} Within this reasoning, the Network Theory of Psychopathology (NTP) has recently been proposed and posits that mental disorders can be understood as systems of interdependent symptoms – so-called symptom-networks.\cite{Robinaugh.2020} From NTP, it follows that we should not aim to differentiate patients and controls per se, but focus on monitoring the neurobiological changes associated with 1) the risk to develop pathological symptom network dynamics and 2) symptom dynamics over longer periods of time. While promising, NTP has not been formalized and the predictive power of the approach remains in question.\cite{Robinaugh.2020, Forbes.2019}  To this end, especially outcome-based, longitudinal research designs are key to advancing our understanding of causal mechanisms with direct impact on the clinical practice.\cite{Opel.2019, Zaremba.2018}  More ecologically valid and easy to administer symptom measurements, e.g., via smartphone applications might aid this endeavor.\cite{Goltermann.2021}  \newline
From these considerations, we derive three recommendations for future research: First, we urge all researchers to clearly communicate the relevance of their finding by reporting effect sizes, predictive utility and distributional overlap in addition to p-values. This will not only enable a more complete assessment of the impact of the results but raise awareness that even highly significant effects can arise in the presence of large distributional overlap and with minimal predictive utility – especially in large-scale studies. Thus, it will help other researchers to judge claims regarding depressive biomarker potential and possible clinical impact of the findings.\cite{Abi-Dargham.2016} If biomarker potential cannot be demonstrated, researchers should precisely state in what way a significant effect advances the development of a quantitative neurobiological theory of depression. This way, even seemingly small effects might be able to explain a specific mechanism or causal relationship within a theoretical framework. Yet, given the lack of quantitative theory in psychiatry and clinical neuroscience, both cases have been deplorably rare in our field.\cite{Fried.2021, Fried.2021nx4} \newline
Second, the community should prioritize more comprehensive phenotyping. This includes deep phenotyping of existing cohorts, the systematic assessment of novel digital phenotypes, as well as longitudinal assessments of symptom dynamics and life events. Ecological validity and minimal interference with patients are two key challenges in this regard. Specifically, large consortia should not only collect larger case-control samples but rather focus on harmonized, clinically relevant outcome variables that allow the testing of informative hypotheses. \newline
Third, while we have focused exclusively on univariate measures in this study, advanced statistical models should be used to address the major issue of poor predictive performance. Within this methodological direction, machine learning approaches are increasingly used to investigate multivariate patterns of deviations and map high-dimensional biological information to complex phenotypes.\cite{Hahn.2017, Winter.2021}  They allow for a seamless integration of multiple modalities – such that overlap of distributions is reduced. Although these methods can also be useful in the context of advancing or falsifying theories, this clear shift from explanation to prediction is more likely to have a direct impact on clinical practice in the short term.\cite{Paulus.2015, Gabrieli.2015}

\subsection*{Conclusion}
In this study, we show that even in a large, fully harmonized study, healthy and depressive participants are remarkably similar on the group level and virtually indistinguishable on the single-subject level across a comprehensive set of neuroimaging modalities. Distributional overlap between healthy controls and patients is dominant, questioning the added informational value these univariate differences provide with regard to theoretical advances or predictive utility. Likewise, increasing phenotypical, biological and methodological homogeneity by running separate analyses for depressive subgroups, individual research sites or gender did not change the overall result. We thus conclude that the small effects of case-control neuroimaging studies of depression published recently are most likely not due to small sample size or heterogeneous study protocols but can be attributed to the striking univariate neurobiological similarity of controls and depressive patients in all neuroimaging modalities commonly used today. Conceptually, we conclude that the phenomenological, descriptive case-control studies which have dominated the last two decades in psychiatric neuroimaging and genetics failed to identify substantial, clinically relevant biological differences between MDD patients and healthy controls. Thus, we urge researchers and funding agencies to go beyond univariate analyses and foster 1) the development of quantitative, theory-driven research as done, e.g., in computational psychiatry, 2) predictive multivariate methods with a clear focus on maximum predictive power and replicability, 3) research into novel measurement approaches, and 4) in-depth phenotyping including longitudinal assessment and digital phenotyping. Future studies will have to investigate whether this can improve clinical utility and theoretical relevance of neurobiological data in mental health. 

\section*{Role of the funding source}
The funders had no role in designing the study; collection, management, analysis, or interpretation of data; writing of the report; nor the decision to submit the report for publication. The authors had the final responsibility for the decision to submit for publication.

\begin{acknowledgements}
This work was funded by the German Research Foundation (DFG grants HA7070/2-2, HA7070/3, HA7070/4 to TH), the Interdisciplinary Center for Clinical Research (IZKF) of the medical faculty of Münster (grants Dan3/012/17 to UD and MzH 3/020/20 to TH) and the Innovative Medizinische Forschung (IMF) of the medical faculty of Münster (grant OP112108 to NO).
The MACS dataset used in this work is part of the German multicenter consortium “Neurobiology of Affective Disorders. A translational perspective on brain structure and function“, funded by the German Research Foundation (Deutsche Forschungsgemeinschaft DFG; Forschungsgruppe/Research Unit FOR2107). Principal investigators (PIs) with respective areas of responsibility in the FOR2107 consortium are: Work Package WP1, FOR2107/MACS cohort and brainimaging: Tilo Kircher (speaker FOR2107; DFG grant numbers KI 588/14-1, KI 588/14-2), Udo Dannlowski (co-speaker FOR2107; DA 1151/5-1, DA 1151/5-2), Axel Krug (KR 3822/5-1, KR 3822/7-2), Igor Nenadic (NE 2254/1-2), Carsten Konrad (KO 4291/3-1). WP2, animal phenotyping: Markus Wöhr (WO 1732/4-1, WO 1732/4-2), Rainer Schwarting (SCHW 559/14-1, SCHW 559/14-2). WP3, miRNA: Gerhard Schratt (SCHR 1136/3-1, 1136/3-2). WP4, immunology, mitochondriae: Judith Alferink (AL 1145/5-2), Carsten Culmsee (CU 43/9-1, CU 43/9-2), Holger Garn (GA 545/5-1, GA 545/7-2). WP5, genetics: Marcella Rietschel (RI 908/11-1, RI 908/11-2), Markus Nöthen (NO 246/10-1, NO 246/10-2), Stephanie Witt (WI 3439/3-1, WI 3439/3-2). WP6, multi method data analytics: Andreas Jansen (JA 1890/7-1, JA 1890/7-2), Tim Hahn (HA 7070/2-2), Bertram Müller-Myhsok (MU1315/8-2), Astrid Dempfle (DE 1614/3-1, DE 1614/3-2). CP1, biobank: Petra Pfefferle (PF 784/1-1, PF 784/1-2), Harald Renz (RE 737/20-1, 737/20-2). CP2, administration. Tilo Kircher (KI 588/15-1, KI 588/17-1), Udo Dannlowski (DA 1151/6-1), Carsten Konrad (KO 4291/4-1). Data access and responsibility: All PIs take responsibility for the integrity of the respective study data and their components. All authors and coauthors had full access to all study data. The FOR2107 cohort project (WP1) was approved by the Ethics Committees of the Medical Faculties, University of Marburg (AZ: 07/14) and University of Münster (AZ: 2014-422-b-S).

\end{acknowledgements}

\begin{interests}
The authors declare no competing financial interests.
\end{interests}

\section*{Bibliography}
\bibliographystyle{zHenriquesLab-StyleBib}
\bibliography{bibliography}